\documentclass[A4paper, 11pt]{article}
\usepackage{amsfonts}
\usepackage{dsfont}
\usepackage{mathrsfs}
\usepackage{amssymb}
\usepackage{bbm}
\usepackage{epsfig}
\usepackage{amsmath}
\usepackage{appendix}
\usepackage{float}
\usepackage[english]{babel}

\def\Reals{\mathop{\hbox{\mit I\kern-.2em R}}\nolimits}

\def\Complexes{{\hbox{\mit C\kern-.46em
            \vrule depth 0ex height 1.4ex width .05em\kern.41em}}}

\newtheorem{thm}{Theorem}
\newtheorem{defn}{Definition}

\newtheorem{lem}{Lemma}
\newtheorem{remark}{Remark}
\newtheorem{prop}{Proposition}

\title{\bf  Randomized Gossip Algorithm with\\ Unreliable Communication\footnote{This work has been supported in part
by the Knut and Alice Wallenberg Foundation, the Swedish Research
Council and  KTH SRA TNG.}}
\date{}

\author{Guodong Shi,  Mikael Johansson and Karl Henrik Johansson\thanks{The authors are with ACCESS Linnaeus Centre, School of Electrical Engineering,
Royal Institute of Technology, Stockholm 10044, Sweden.
       Email: {\tt\small $\{$guodongs, mikaelj, kallej$\}$@kth.se}}}

\begin{document}

\maketitle
%%%%%%%%%%%%%%%%%%%%%%%%%%%%%%%%%%%%%%%%%%%%%%%%%%%%%%%%%%%%%%%%%%%%%%%%%%%%%%%%%%%%
\begin{abstract}
In this paper,  we study an asynchronous randomized gossip algorithm under unreliable communication. At each instance, two nodes are selected to meet with a given probability. When nodes meet, two unreliable communication links are established with communication in each direction succeeding with a time-varying probability. It is shown that two particularly interesting cases arise when these communication processes are either perfectly dependent or independent.  Necessary and sufficient conditions on the success probability sequence are proposed to ensure almost sure consensus or $\epsilon$-consensus. Weak connectivity is required when the communication is perfectly dependent, while double connectivity is required when the communication is independent. Moreover, it is proven that with odd number of nodes, average preserving turns from  almost forever (with probability one for all initial conditions) for perfectly dependent communication, to almost never (with probability zero for almost all initial conditions) for the independent case. This average preserving property does not hold true for general number of nodes. These results indicate the fundamental role the node interactions have in randomized gossip algorithms.
\end{abstract}


%%%%%%%%%%%%%%%%%%%%%%%%%%%%%%%%%%%%%%%%%%%%%%%%%%%%%%%%%%%%%%%%%%%%%%%%%%%%%%%%%%%%%
\section{Introduction}
Peer-to-peer networks, sensor networks and social networks constitute a new generation of engineering systems that has grown tremendously in importance over the last decade. The absence of a central decision-making entity, the unprecedented number of interacting nodes, the time-varying topology of node interactions, and the unreliability of nodes are key challenges for the analysis and design of these systems. Gossiping algorithms, in which each node exchanges data and decisions with at most one neighboring node in each time slot, have proven to be a robust and efficient way to structure distributed computation and information processing over such networks  \cite{gossip1, gossip2, boyd1, shah}. A limitation of the current literature is that while it allows node interactions to be random, it assumes that the probabilities that two specific nodes interact is constant in time, and that when two nodes interact, both nodes completely and correctly execute the proposed algorithm. However, unreliable communication in wireless networks, and asymmetry of trust in social networks challenge such assumptions. This paper develops a framework for analysis of gossip algorithms that separates the random process for node interactions with the random process for successful information exchange and algorithm execution, and allows time-varying success probabilities for these operations. Necessary and sufficient conditions for the underlying graph structure and probability sequences are developed to ensure a.s. or $\epsilon$-consensus under both perfectly dependent communication, in which the bidirectional message exchange between nodes either succeeds of fails, and independent communication, in which the success of communication in each direction is independent of the outcome in the other direction.

Gossip algorithms for distributed averaging arise in many applications, such as load balancing in parallel computing \cite{cs2,cs3}, coordination of autonomous agents  \cite{jad03}, distributed estimation \cite{moura1} and analysis of opinion dynamics \cite{daron}. A central problem here is to analyze if a given gossip algorithm converges to consensus, and to determine the rate of convergence of the consensus process: Karp et al. \cite{gossip2} derived a general lower bound for synchronous gossiping; Kempe et al. \cite{gossip1} proposed a randomized gossiping algorithm on complete graphs and determined the order of its convergence rate; and Boyd et al. \cite{boyd} established both lower and upper bounds for the convergence time of synchronous and asynchronous randomized gossiping algorithms, and developed algorithms for optimizing parameters to obtain fast consensus. In \cite{daron}, a gossip algorithm was used to describe the spread of misinformation in social networks, where the state of each node was viewed as its belief and the randomized gossip algorithm characterized the dynamics of the belief evolution. A detailed introduction to gossip algorithms can be found in \cite{shah}.

More generally, consensus problems on graphs have been investigated by researchers from many disciplines, including computer science \cite{cs2,cs3}, engineering \cite{caoming3, roy, mar,jad03,tsi, ren} and social sciences \cite{degroot,como}. Deterministic consensus algorithms have been studied extensively both for time-invariant and time-varying communication graphs. Efforts have typically been devoted to finding connectivity conditions which ensure convergence to consensus for the network model under consideration \cite{tsi, jad03, tsi2, mor, mar, ren, caoming1, caoming2}. Randomized consensus algorithms, motivated by the stochastic information flow over networks, have also been considered \cite{hatano, wu, fagnani1, fagnani2, jad2, moura1}. Many sufficient and/or necessary conditions have been established that guarantee a global consensus with probability one \cite{hatano, wu, fagnani1, fagnani2, jad2, aysal}  under different settings for the randomization of the communication graph and convergence rates have also, in some cases, been established \cite{fagnani1, fagnani2, bamieh}.

Most of the existing work on randomized consensus algorithms have focused on the influence of randomness in the underlying communication graph on the agreement seeking process. In \cite{boyd,f-z,daron}, the communication graph was determined by a stochastic matrix whose elements specify the probability that each pair of nodes is selected to execute the gossip algorithm with each other in a given time instant. In \cite{markov}, the communication graph was described by a finite-state Markov chain where each state corresponds to a specific realization of the communication graph, and conditions for guaranteeing almost sure consensus were given. In \cite{hatano}, the authors studied linear consensus dynamics with communications graphs given by a sequence of independent and identically distributed (i.i.d.) Erd\"{o}s-R\'{e}nyi random graphs; the analysis was later extended to directed Erd\"{o}s-R\'{e}nyi graphs in \cite{wu}. Mean-square performance for consensus algorithms over i.i.d. random graphs was studied in \cite{fagnani1} and the influence of random packet drops was investigated in \cite{fagnani1,bamieh}. Distributed consensus over random graphs with graph independence was also studied in \cite{jad2,moura1}.  In all these works, the randomness in the underlying communication graph describes the randomness in the process of initiating interactions. Even if two nodes meet with each other, the information exchange could still fail, or be only partially executed (e.g., only by one of the two nodes) due to unreliable communication between nodes and the asymmetry of trust.

This paper studies an asynchronous randomized gossip algorithm under unreliable  communication. A key feature of the proposed analysis framework is that it separates the random process for \emph{initiating interactions} between node pairs from the random process for \emph{successful communication} between two interacting nodes, and that it considers time-varying success probabilities for inter-node communication. Both perfectly dependent  communication, in which the bi-directional message exchange between two nodes either succeeds or fails, and independent communication, in which the success of communication in each direction is independent of the outcome in the other direction, are considered. The perfectly dependent and independent node communication reflects directly the symmetry of the gossiping algorithms \cite{f-z}. We derive necessary and sufficient conditions on the underlying graph structure and the communication success probability sequence to ensure a.s. consensus or $\epsilon$-consensus under perfectly dependent and independent communication, respectively. We investigate the impact of communication dependence on the behavior of the algorithm, and show how the underlying requirement for reaching a.s. consensus goes from weak connectivity in the perfectly dependent case to double connectivity in the independent case, while the probability of preserving the average decreases from one for perfectly dependent communication to zero for the independent case when the number of nodes is odd.

The rest of the paper is organized as follows. In Section 2, some preliminary concepts are introduced. We present the network model, the randomized gossip algorithm, and the standing assumptions  in Section 3. Then main results   for perfectly dependent  and independent communication are given   in Sections 4 and 5, respectively.   Finally, concluding remarks are given in Section 6.

\section{Preliminaries}
In this section, we recall  some  basic definitions from graph theory \cite{god}, stochastic matrices \cite{lat} and Bernoulli trials \cite{bert}.

\subsection{Directed Graphs}A directed graph (digraph) $\mathcal
{G}=(\mathcal {V}, \mathcal {E})$ consists of a finite set
$\mathcal{V}$ of nodes and an arc set
$\mathcal {E}\subseteq \mathcal{V}\times\mathcal{V}$.  An element $e=(i,j)\in\mathcal {E}$ is called an
{\it arc}  from node $i\in \mathcal{V}$  to $j\in\mathcal{V}$. If the
arcs are pairwise distinct in an alternating sequence
$ v_{0}e_{1}v_1e_{2}v_{2}\dots e_{k}v_{k}$ of nodes $v_{i}\in\mathcal{V}$ and
arcs $e_{i}=(v_{i-1},v_{i})\in\mathcal {E}$ for $i=1,2,\dots,k$,
the sequence  is called a (directed) {\it  path} with {\it length} $k$. A path with no repeated nodes is called a {\it simple path}. A path from $i$ to
$j$ is denoted as $i \rightarrow j$, and the length of $i \rightarrow j$ is denoted as $|i \rightarrow j|$.  If there exists a path from node $i$ to node $j$,
then node $j$ is said to be reachable from node $i$.  Each node is thought to be reachable by itself. A node $v$ from which any other node is
reachable  is called a {\it center} (or a {\it root}) of $\mathcal {G}$. A digraph $\mathcal
{G}$ is said to be {\it strongly connected}  if it contains path $i \rightarrow j$ and $j \rightarrow i$ for every pair of nodes $i$ and $j$, and {\it
quasi-strongly connected} if $\mathcal {G}$ has a center
\cite{ber}.

The {\it converse graph}, $\mathcal{G}^T$ of a digraph $\mathcal
{G}=(\mathcal {V}, \mathcal {E})$, is defined as the graph obtained by reversing the orientation of all arcs in $ \mathcal {E}$. The {\it distance} from $i$ to $j$ in a digraph $\mathcal{G}$, $d(i,j)$,   is the length of a shortest simple path $i \rightarrow j$ if $j$ is reachable from $i$, and the {\it diameter} of $\mathcal
{G}$ is $\rm{diam}(\mathcal{G})$$=\max\{d(i,j)|i,j \in\mathcal
{V},\ j\mbox{ is reachable from}\ i\}$.

The
union of two digraphs with the same node set $\mathcal {G}_1=(\mathcal {V},\mathcal {E}_1)$ and $\mathcal
{G}_2=(\mathcal {V},\mathcal {E}_2)$ is defined as $\mathcal {G}_1\cup\mathcal
{G}_2=(\mathcal {V},\mathcal {E}_1\cup\mathcal {E}_2)$; we denote $\mathcal {G}_1 \subseteq \mathcal
{G}_2$ if $\mathcal {E}_1 \subseteq \mathcal
{E}_2$. A digraph $\mathcal {G}$ is said to be bidirectional if for every two nodes $i$ and $j$, $(i,j)\in \mathcal{E}$  if and only if $(j,i)\in \mathcal{E}$ .  A bidirectional graph $\mathcal{G}$ is said to be {\it connected} if there is a path between any two nodes. A digraph $\mathcal
{G}$ is said to be {\it weakly connected} if it is connected as a bidirectional graph when all the arc directions are ignored. Strongly or quasi-strongly connected digraphs are hence always weakly connected.
\subsection{Stochastic Matrices}
A finite square matrix $M=[m_{ij}]\in\mathds{R}^{n\times n}$ is called {\em stochastic} if $m_{ij}\geq 0$ for all $i,j$ and $\sum_j m_{ij}=1$ for all $i$. For a stochastic matrix $M$, introduce
\begin{equation}\label{origin}
\delta (M)=\max_j \max_{\alpha, \beta}|m_{\alpha j}-m_{\beta j}|, \quad \lambda (M)=1-\min_{\alpha, \beta}\sum_j \min \{m_{\alpha j},m_{\beta j}\}.
\end{equation}
If $\lambda (M)<1$ we call $M$ a {\em scrambling} matrix. The following lemma  can be found in \cite{haj}.

\begin{lem}\label{lem0}For any $k\geq 1$ stochastic matrices $M_1,\dots,M_k$,
\begin{equation}
\delta (M_k\cdots M_2 M_1)\leq \prod_{i=1}^{k}\lambda (M_i).
\end{equation}
\end{lem}

A stochastic matrix $M=[m_{ij}]\in\mathds{R}^{n\times n}$ is called {\em doubly stochastic} if also $M^T$ is  stochastic.

Let $P=[p_{ij}]\in\mathds{R}^{n\times n}$ be a matrix with nonnegative entries. We can associate a unique digraph  $\mathcal{G}_P=\{\mathcal{V},\mathcal{E}_P\}$ with $P$ on node set $\mathcal{V}=\{1,\dots,n\}$ such that $(j,i)\in\mathcal{E}_P$ if and only if $p_{ij}>0$. We call $\mathcal{G}_P$ the {\em induced graph} of $P$.
\subsection{Bernoulli Trials}
 A sequence of independently distributed Bernoulli trials is a finite or infinite sequence of independent random variables $\mathfrak{B}_0, \mathfrak{B}_1, \mathfrak{B}_2, \dots$, such that
\begin{itemize}
\item[(i)] For each $k\geq 0$, $\mathfrak{B}_k$ equals  either $0$ or $1$;
\item [(ii)] For each $k\geq 0$, the probability that $\mathfrak{B}_k=1$ is $p_k$.
\end{itemize}
We call $p_k$ the success probability for time $k$. The sequence of integers
\begin{align}
 0\leq \zeta_1<\zeta_2<\dots :\ \ \    \mathfrak{B}_{\zeta_m} = 1
\end{align}
is called the {\em Bernoulli (success)} sequence associated with the sequence of Bernoulli trials with $\zeta_m$ marking the time of the $m$'th success.

\section{Problem Definition}
In this section, we present the considered network model and define the problem of interest.
\subsection{Node Pair Selection Process}
Consider a network  with node set $\mathcal{V}=\{1,\dots,n\}$ ($n\geq3$).  Let the digraph $\mathcal {G}_0=(\mathcal{V},\mathcal{E}_0)$ denote the {\it underlying}  graph of the considered network. The underlying graph indicates  potential interactions between nodes. We use the asynchronous time model introduced in \cite{boyd} to describe node  interactions. Each node meets other nodes at independent time instances  defined  by a  rate-one Poisson process. This is to say, the inter-meeting times at each node follows a rate-one  exponential distribution. Without loss of generality, we can assume that  at most one node is active at any given instance. Let $x_i(k)\in\mathds{R}$ denote the state (value) of node $i$ at the $k$'th meeting slot among all the nodes.

 Node interactions are characterized by an $n\times n$ matrix $A=[a_{ij}]$, where $a_{ij}\geq0$ for all $i,j=1,\dots,n$ and $a_{ij}>0$ if and only if $(j,i)\in \mathcal{E}_0$. We assume $A$ is a stochastic matrix. The meeting process is defined as follows.

\vspace{3mm}
The node pair selection process for the gossip algorithm is defined as follows.

\begin{defn} (Node Pair Selection Process)
At each time $k\geq0$,
\begin{itemize}
\item[(i)]  A node $i\in\mathcal{V}$ is drawn    with probability $1/n$;
  \item[(ii)] Node $i$ picks the pair $(i,j)$ with probability $a_{ij}$.
\end{itemize}
\end{defn}

Note that, by the definition of the node pair selection process, the underlying graph $\mathcal{G}_0$ is actually the same as $\mathcal{G}_A$, the induced graph of the node pair selection matrix $A$.  For $\mathcal{G}_0$, we use the following assumption.

\vspace{3mm}

  \noindent {\bf A1.} {\em (Weak Connectivity)} The underlying graph $\mathcal {G}_0$ is weakly connected.

\begin{remark}
Node that, node pairs $(i,j)$ and $(j,i)$ have different meaning according to the node pair selection process. When $(i,j)$ is selected, node $i$ is the node first picked and then $i$ picks $j$. While pair $(j,i)$ is selected means  $j$ is first picked who picks $i$ later.
\end{remark}

  \begin{remark} The node pair selection matrix $A$ being a stochastic matrix has  natural meaning that node $i$'s decisions form a well-defined probability space when it is selected at time $k$. However, it is not essential for the following discussions in the sense that all the results still stand even this assumption is replaced by the condition that each row sum of $A$ is no larger than one. The same assumption is made in \cite{boyd, daron}.
  \end{remark}

\begin{remark}\label{rem1}
 Let $\mathcal{G}_{A+A^T}$ denote the induced graph of matrix $A+A^T$. Apparently $A+A^T$ is symmetric, and thus $\mathcal{G}_{A+A^T}$ is a bidirectional graph. It is not hard to see that $\mathcal {G}_0$ is weakly connected if and only if $\mathcal{G}_{A+A^T}$ is a connected bidirectional graph.
\end{remark}
\begin{remark}
In the standing assumption of \cite{boyd}, the matrix $A$ is supposed to have its largest eigenvalue equal to $1$ and all other $n-1$ eigenvalues strictly less than $1$ in magnitude. This condition is equivalent with that  $\mathcal{G}_0$ is quasi-strongly connected \cite{ren, caoming1}. On the other hand, in \cite{daron}, $\mathcal {G}_0$ is assumed to be strongly connected. Therefore, Assumption 1 is a weaker assumption, compared to the one in \cite{boyd, daron}.
\end{remark}
\begin{remark}\label{rem4}
In order to guarantee  convergence for the  gossip algorithm discussed below, A1 cannot be further weakened based on the following argument. Let us just assume A1 does not hold true. Then there will be two disjoint node sets $\mathcal{V}_1, \mathcal{V}_2\subset \mathcal{V}$ such that there is no link  connecting the two sets. As a result, nodes in $\mathcal{V}_i, i=1,2$ can only communicate with nodes belonging to  the same subset. Therefore, the network is essentially divided into two isolated parts, and a convergence for the whole network is thus impossible.
\end{remark}
\subsection{Node Communication Process}
When pair $(i,j)$ is selected, both nodes try to set their states equal to the average of their current states. To this end, two communication links with opposite directions are established between the two nodes.

%\begin{figure}[H]
%\centerline{\epsfig{figure=channel.eps, width=0.50\linewidth=0.25}}
%\caption{Unreliable node communication. }\label{sss}
%\end{figure}

Let $\{P_k^+\}_0^\infty$ and $\{P_k^-\}_0^\infty$ be two deterministic sequences with $0\leq P_k^+, P_k^-\leq1$ for all $k$. The node communication process is defined as follows.

\begin{defn}{(Node Communication Process)} Independent with time, node states, and the node pair selection process,
\begin{itemize}
\item[(i)] $\mathbf{P} \big(\mathbb{E}^+_k \big)=P_k^+$ with   $\mathbb{E}^+_k=\big\{$node $i$ receives $x_j(k)$ when $(i,j)$ is selected at time $k\big\}$;
\item[(ii)] $\mathbf{P} \big(\mathbb{E}^-_k\big)=P_k^-$ with   $\mathbb{E}^-_k=\big\{$node $j$ receives $x_i(k)$ when $(i,j)$ is selected at time $k\big\}$.
\end{itemize}
\end{defn}

If a  node fails to receive the value of the other node, it will keep its current state. Note that we do not in general impose independence between $i$ receiving $x_j(k)$ and $j$ receiving $x_i(k)$ when pair $(i,j)$ is selected.  In fact, we will study how such potential dependence in the communication process influences the convergence of the gossip algorithm.

\begin{remark}
A randomized gossip algorithm  can also be viewed as belief propagation  in a social network, where $x_i(k)$ represents the belief of node $i$. Then the communication process naturally  captures the loss of 'trust' when two nodes meet and exchange opinions \cite{daron,como}. Therefore, from a social network viewpoint, the discussion in this paper on the convergence property of the gossip algorithm establishes the influence of missing `trust' in belief agreement.
\end{remark}

\subsection{Problem}
Let the initial condition be $x^0=x(k_0)=(x_1(k_0)\dots x_n(k_0))^T\in \mathds{R}^{n}$, where $k_0\geq 0$ is an arbitrary integer. According to the node pair selection process and the node communication process, the iteration of the gossip algorithm can be expressed as: for $k\geq k_0$,
\begin{equation}\label{00}
	x_i(k+1) =
	\begin{cases}
		\frac{1}{2}x_{i}(k)+\frac{1}{2}x_{j}(k), & \text{if $\mathcal{M}^{\langle i,j\rangle}$ happens}\\
		\ \ x_i(k), & \text{otherwise,}\\
	\end{cases}
\end{equation}
where $$
\mbox{$\mathcal{M}_k^{\langle i,j\rangle}\doteq\Big\{(i,j)$ is selected or $(j,i)$ is selected,  and $i$ receives $x_{j}(k)$ at time $k\Big\}$}
$$
denotes the event that node $i$ successfully updates at time $k$. According to the definitions above, we have
$$
\mathbf{P}\Big(\mathcal{M}_k^{\langle i,j\rangle}\Big)=\frac{a_{ij}}{n}P^+_k+\frac{a_{ji}}{n}P^-_k; \quad \quad \mathbf{P}\Big(\mathcal{M}_k^{\langle j,i\rangle}\Big)=\frac{a_{ji}}{n}P^+_k+\frac{a_{ij}}{n}P^-_k.$$
Therefore, the two events, $\mathcal{M}_k^{\langle i,j\rangle}$ and $\mathcal{M}_k^{\langle j,i\rangle}$, are not necessarily symmetric in their probabilities, due to the potential asymmetry of the meeting matrix $A$.

In this paper, we study the convergence of the  randomized gossip consensus algorithm and  the time it takes for the network to reach a consensus. Let $$
x(k;k_0,x^0)=\Big(x_1\big(k;k_0,x_1(k_0)\big)\dots x_n\big(k;k_0,x_n(k_0)\big)\Big)^T\in \mathds{R}^{n}
$$ be the random process driven by the randomized algorithm (\ref{00}). When it is clear from the context, we will identify $x(k;k_0,x^0)$ with $x(k)$.

Denote
$$
 H(k)\doteq\max_{i=1,\dots,n}x_{i}(k), \quad h(k)\doteq\min_{i=1,\dots,n}x_{i}(k)
$$
as the maximum and minimum  states among all nodes, respectively, and define $\mathcal{H}(k)\doteq H(k)-h(k)$ as the consensus metric. We introduce the following definition.
\begin{defn} (i) A global a.s. {\it consensus}  is achieved  if
\begin{equation}\label{4}
\mathbf{P}(\lim_{k\rightarrow \infty} \mathcal{H}(k)=0)=1
\end{equation}
for any initial condition $x^0\in \mathds{R}^{n}$.

(ii) Let the $\epsilon$-computation time be
\begin{equation}\label{4}
T_{\rm com}(\epsilon)\doteq\sup_{x(k_0)} \inf \Big\{ k-k_0:\ \  \mathbf{P}\Big(\frac{\mathcal{H}(k)}{\mathcal{H}(k_0)}\geq \epsilon\Big)\leq \epsilon\Big\}.
\end{equation}
Then a global a.s. $\epsilon$-consensus  is achieved  if
\begin{align}
T_{\rm com}(\epsilon)=O(\log \epsilon^{-1})
\end{align}
where by definition $f(\epsilon)=O\big(g(\epsilon)\big)$ means that $\limsup_{\epsilon\rightarrow0}{f(\epsilon)}/{g(\epsilon)}<\infty$ is a nonzero  constant.
\end{defn}
\begin{remark}
A global a.s. only requires that $\mathcal {H}(t)$ will converge to zero with probability one. If it is further required that the convergence speed is sufficiently  fast,  we use global a.s. $\epsilon$-agreement. The $\epsilon$-computation $T_{\rm com}(\epsilon)$ is essentially equivalent  with the definition of $\epsilon$-averaging time in \cite{boyd}, which is a probabilistic version of similar concepts  used to characterize the convergence rate of deterministic consensus algorithms in the literature, e.g., \cite{tsi2}.
\end{remark}

Recall that until now,  when $i$ is selected to meet node $j$ at time $k$, no assumption has been made on the dependence between the communication from $i$ to $j$, and the  one from $j$ to $i$. In the following two sections, we will discuss the convergence of the considered randomized gossip algorithm  with perfectly dependent and independent communication, respectively. We will show that the dependence in the node communication plays a critical role in determining the behavior of the gossip algorithm.

\section{Convergence under Perfectly Dependent Communication}
In this section, we study the case when the  communication between nodes $i$ and $j$ is perfectly dependent,  as described in the following assumption.

\vspace{3mm}

 \noindent {\bf A2.} {\em (Perfectly Dependent Communication)} The events  $\mathbb{E}^+_k=\mathbb{E}^-_k$ except for a set with probability zero for all $k$.

\vspace{3mm}

 Note that A2 is equivalent to assuming  that $\mathbf{P}(\mathbb{E}^+_k|\mathbb{E}^-_k)=\mathbf{P}(\mathbb{E}^-_k|\mathbb{E}^+_k)=1$. Hence, we have $P^+_k=P^-_k$ and at each instant, with probability $P_k\doteq P^+_k=P^-_k$, both $\mathbb{E}^+_k$ and $\mathbb{E}^-_k$ occur, and with probability $1-P_k$ they both fail.  With A2, the gossip algorithm can be expressed as
 \begin{align}\label{1}
 x(k+1)=W(k)x(k),
 \end{align}
where $W(k)$ is the random matrix satisfying
\begin{align}\label{2}
\mathbf{P}\Big(W(k)=W_{\langle ij\rangle}\doteq I-\frac{(e_i-e_j)(e_i-e_j)^T}{2}\Big)=\frac{a_{ij}+a_{ji}}{n}{P}_k,\ \ \ \  i\neq j
\end{align}
 with $e_m=(0 \dots 0\  1\  0 \dots 0)^T$ denoting the $n\times1$ unit vector whose $m$'th component is $1$. Moreover, $\mathbf{P}\big(W(k)= W_{\langle ii\rangle}=I\big)=1-\sum_{i>j }\frac{a_{ij}+a_{ji}}{n}{P}_k$.

 The main result on a.s. consensus for the considered  gossip algorithm under perfectly dependent  communication is stated as follows.
\begin{thm}\label{thm1}
Suppose A1 (Weak Connectivity) and A2 (Perfectly Dependent Communication) hold. Global a.s. {\it consensus} is achieved  if and only if $\sum_{k=0}^\infty {P}_k=\infty$.
\end{thm}

Denote $D=\mbox{diag}(d_1 \dots d_n)$ with $d_i=\sum_{j=1}^n (a_{ij}+a_{ji})$. For a.s. $\epsilon$-consensus, we have the following conclusion.
\begin{thm}\label{thm2}
Suppose A1 (Weak Connectivity) and A2 (Perfectly Dependent Communication) hold. Global a.s. {\it $\epsilon$-consensus}  is achieved  if and only if there exist a constant $p_\ast>0$ and an integer $T_\ast\geq1$ such that $\sum_{k=m}^{m+T_\ast-1} {P}_k\geq p_\ast$ for all $m\geq0$. In fact,   we have
\begin{align}
T_{\rm com}(\epsilon)\leq {3}\Big[{\log \big(1-\frac{\lambda_2^\ast p_\ast}{2nT_\ast}\big)^{-1}}\Big]^{-1}\log \epsilon^{-1}+O(1),
\end{align}
where $\lambda_2^\ast$ is the second smallest eigenvalue of $D-(A+A^T)$.
\end{thm}

Theorem \ref{thm1} indicates that $\sum_{k=0}^\infty {P}_k=\infty$ is actually a threshold for the gossip algorithm to reach a.s. consensus. For $\epsilon$-consensus,  Theorem \ref{thm2}  implies that $\sum_{k=0}^m {P}_k=O(m)$ is the  threshold condition, which actually requires that $\sum_{k=0}^m {P}_k$ grows linearly as a function of $m$.
\begin{remark}
Theorems \ref{thm1} and \ref{thm2} rely on the fact that there are at least three nodes in the network. If the network contains only two nodes, then both Theorems \ref{thm1} and \ref{thm2} no longer hold true. This phenomenon  is interesting since many consensus results in the literature are independent of  the number of nodes, e.g., \cite{jad03,ren,tsi,caoming1,caoming2,caoming3}.
\end{remark}

Let the random variable $\xi(k_0,x^0)$ denote the consensus limit (supposed to exist), i.e.,
\begin{align}
\lim_{k\rightarrow \infty}x_i(k)=\xi, \ \ a.s.\ \ \quad i=1,\dots, n.
\end{align}
Denote $x_{\rm ave}=\sum_{i=1}^n{x_i(k_0)}/n$ be the average of the initial values. Then the following conclusion holds showing that the average is preserved almost surely with perfectly dependent communication.

\begin{thm}\label{thm3}
Suppose A1 (Weak Connectivity) and A2 (Perfectly Dependent Communication) hold.  Then for all initial conditions $x^0=x(k_0)\in\mathds{R}^n$, we have
\begin{align}\label{21}
\mathbf{P}\Big(\sum_{i=1}^n x_{i}(k)=n x_{\rm ave},\ \  k\geq k_0\Big)=1.
 \end{align}
 Consequently, we have $\mathbf{P}\big(\xi=x_{\rm ave}\big)=1$  whenever the consensus limit exists.
\end{thm}

In the following two subsections, we will present the proof of Theorems \ref{thm1} and \ref{thm2}, respectively. Theorem \ref{thm3} follows  from the proof Theorem \ref{thm1}.

The upcoming analysis  relies on the following well-known lemmas.
\begin{lem}\label{lem1}
Suppose $0\leq b_k<1$ for all $k$. Then $\sum_{k=0}^\infty b_k=\infty$ if and only if $\prod_{k=0}^{\infty}(1-b_k)=0$.
\end{lem}
\begin{lem}\label{lem2}
$\log(1-t)\geq -2t$ for all $0\leq t\leq {1}/{2}$.
\end{lem}
\subsection{Proof of Theorem \ref{thm1}}
(Sufficiency.) This part of the proof is based on a similar argument as is used in \cite{boyd}. Define $L(k)=\sum_{i=1}^n |x_i(k)-x_{\rm ave}|^2$, where $|\cdot|$ represents the Euclidean norm of a vector or the absolute value of a scalar.

It is easy to verify  for every possible sample and fixed instant $k$ that $W_{\langle ij\rangle}$ of the random matrix $W(k)$ defined in (\ref{1}) and (\ref{2}) fulfills
 \begin{itemize}
 \item[(i).] $W_{\langle ij\rangle}$ is a  doubly stochastic matrix, i.e., $W_{\langle ij\rangle}\mathbf{1}=\mathbf{1}$ and $\mathbf{1}^T W_{\langle ij\rangle}=\mathbf{1}^T$;
 \item[(ii).] $W_{\langle ij\rangle}$ is a  projection matrix, i.e., $W_{\langle ij\rangle}=W_{\langle ij\rangle}^T W_{\langle ij\rangle}$.
 \end{itemize}
Therefore, we have
\begin{align}\label{3}
\mathbf{E}\Big(L(k+1)\big|x(k)\Big)&=\mathbf{E}\Big( \big(x(k+1)-x_{\rm ave}\mathbf{1}\big)^T \big(x(k+1)-x_{\rm ave}\mathbf{1}\big)\big|x(k)\Big)\nonumber\\
&=\mathbf{E}\Big( \big(W(k)x(k)-x_{\rm ave}\mathbf{1}\big)^T \big(W(k)x(k)-x_{\rm ave}\mathbf{1}\big)\big|x(k)\Big)\nonumber\\
&=\mathbf{E}\Big( \big(x(k)-x_{\rm ave}\mathbf{1}\big)^TW(k)^TW(k) \big(x(k)-x_{\rm ave}\mathbf{1}\big)\big|x(k)\Big)\nonumber\\
&=\big(x(k)-x_{\rm ave}\mathbf{1}\big)^T \mathbf{E}\big(W(k)^TW(k)\big) \big(x(k)-x_{\rm ave}\mathbf{1}\big)\nonumber\\
&=\big(x(k)-x_{\rm ave}\mathbf{1}\big)^T \mathbf{E}\big(W(k)\big) \big(x(k)-x_{\rm ave}\mathbf{1}\big)
\end{align}

Since $W(k)$ is doubly stochastic, we know  that the sum of the nodes' states, $\sum_{i=1}^n x_i(k)$, is preserved with probability one, and $\mathbf{1}$ is the eigenvector corresponding to eigenvalue $1$ of $\mathbf{E}\big(W(k)\big)$ (Theorem \ref{thm3} therefore holds). Thus, we can conclude from (\ref{3}) that
\begin{align}\label{5}
\mathbf{E}\Big(L(k+1)\big|x(k)\Big)
&\leq \lambda_2 \Big(\mathbf{E}\big(W(k)\big)\Big)\big(x(k)-x_{\rm ave}\mathbf{1}\big)^T  \big(x(k)-x_{\rm ave}\mathbf{1}\big)=\lambda_2 \Big(\mathbf{E}\big(W(k)\big)\Big)L(k),
\end{align}
where $\lambda_2(M)$ for a stochastic matrix $M$ denotes the largest eigenvalue in magnitude except for the eigenvalue at one. Here note that $\mathbf{E}\big(W(k)\big)$ is symmetric and positive semi-definite.

Now according to (\ref{2}), we see that
\begin{align}\label{6}
\mathbf{E}\big(W(k)\big)=I-\frac{{P}_k}{2n}\big(D-(A+A^T)\big).
\end{align}
 Note that $D-(A+A^T)$ is actually the (weighted) Laplacian of the graph $\mathcal{G}_{A+A^T}$. With assumption A1, $\mathcal{G}_{A+A^T}$ is a connected graph (cf., Remark \ref{rem1}),  and therefore, based on the well-known property of Laplacian matrix of connected graphs \cite{god}, we have $\lambda_2^\ast>0$, where $\lambda_2^\ast$ is the second smallest eigenvalue of $D-(A+A^T)$. On the other hand, since $A$ is a stochastic matrix, it is straightforward to see that
 \begin{align}
 \sum_{j=1,j\neq i}a_{ij}+a_{ji}\leq n
 \end{align}
for all $i=1,\dots,n$. According to Gershgorin circle theorem, all the eigenvalues of $D-(A+A^T)$ are bounded by $2n$. Therefore, we conclude from (\ref{6}) that for all $k$,
\begin{align}\label{r1}
\lambda_2 \Big(\mathbf{E}\big(W(k)\big)\Big)= 1-\frac{\lambda_2^\ast}{2n}{P}_k.
\end{align}

 With (\ref{5}) and (\ref{r1}), we obtain
\begin{align}\label{10}
\mathbf{E}\Big(L(k+1)\Big)
\leq \prod_{i=k_0}^{k}\lambda_2 \Big(\mathbf{E}\big(W(i)\big)\Big)L(k_0)=\prod_{i=k_0}^{k}\Big(1-\frac{\lambda_2^\ast}{2n}{P}_i\Big)L(k_0),
\end{align}

Therefore, based on Lemma \ref{lem1} and Fatou's lemma, we have
\begin{align}
\mathbf{E}\Big(\lim_{k\rightarrow\infty}L(k)\Big)\leq\lim_{k\rightarrow\infty}\mathbf{E}\Big(L(k)\Big)=0,
\end{align}
where $\lim_{k\rightarrow\infty}L(k)$ exists simply from the fact that the sequence is non-increasing. This immediately implies
\begin{align}
\mathbf{P}\big(\lim_{k\rightarrow\infty} x_i(k)=x_{\rm ave}\big)=1.
\end{align}
The sufficiency claim of the theorem thus holds.

\vspace{3mm}

\noindent (Necessity.) From the definition of the gossip algorithm, we have
\begin{align}\label{24}
\mathbf{P}\big(x_i(k+1)=x_i(k)\big)&\geq 1- \mathbf{P}\big(\mbox{$i$ receives $x_j(k)$}\big)\cdot\sum_{j=1,\ j\neq i}^n\Big[\mathbf{P}\big(\mbox{pair $(i,j)$ is selected}\big)\nonumber\\
&\ \ \ \ +\mathbf{P}\big(\mbox{pair $(j,i)$ is selected}\big)\Big]\nonumber\\
&=1-{P}_k\sum_{j=1,\ j\neq i}^n\frac{1}{n}\big(a_{ij}+a_{ji}\big)\nonumber\\
&\doteq 1- h_i {P}_k,
\end{align}
where $h_i=\sum_{j=1,\ j\neq i}\frac{1}{n}\big(a_{ij}+a_{ji}\big), i=1,\dots,n$. Noting the fact that
\begin{align}
\sum_{i=1}^n h_i =\sum_{i=1}^n\sum_{j=1,\ j\neq i}^n\frac{1}{n}\big(a_{ij}+a_{ji}\big)= 2-\sum_{i=1}^n a_{ii}\leq2,
\end{align}
there exists at least one node $\alpha_1\in\mathcal{V}$ such that $h_{\alpha_1}<1$ since $n\geq 3$. Moreover, assumption A1 further guarantees that $h_i>0,i=1,\dots,n$, which implies that there exists another node $\alpha_2\in\mathcal{V}$ such that $h_{\alpha_2}<1$.

Therefore,  if $\sum_{k=0}^\infty P_k<\infty$, we have
\begin{align}\label{15}
\mathbf{P}\big(x_{\alpha_i}(k)=x_{\alpha_i}(k_0),k\geq k_0\big)=\prod_{k=k_0}^{\infty} \big(1- h_{\alpha_i} {P}_k\big)\doteq \sigma_{i}>0, \quad i=1,2
\end{align}
based on Lemma \ref{lem1}.  Consequently, choosing $x_{\alpha_1}(k_0)\neq x_{\alpha_2}(k_0)$, consensus will fail with probability $\sigma_{1}\sigma_{2}>0$. This completes the proof.
\subsection{Proof of Theorem \ref{thm2}}
(Sufficiency.) Recall that $\mathcal{H}(k)\doteq \max_{i=1,\dots,n}x_{i}(k)-\min_{i=1,\dots,n}x_{i}(k)$. Suppose nodes $\alpha$ and $\beta$ reach the maximum and minimum values at time $k$, respectively, i.e.,
$$x_\alpha(k)=\max_{i=1,\dots,n}x_{i}(k); \quad x_\beta(k)=\min_{i=1,\dots,n}x_{i}(k).
$$
Then we have
\begin{align}\label{7}
L(k)=\sum_{i=1}^n |x_i(k)-x_{\rm ave}|^2&\geq |x_\alpha(k)-x_{\rm ave}|^2+|x_\beta(k)-x_{\rm ave}|^2\nonumber\\
&\geq \frac{1}{2}|x_\alpha(k)-x_\beta(k)|^2\nonumber\\
&=\frac{1}{2}\mathcal{H}^2(k).
\end{align}
On the other hand,
\begin{align}\label{8}
L(k_0)=\sum_{i=1}^n |x_i(k_0)-x_{\rm ave}|^2&\leq \frac{1}{n^2}\Big(\sum_{i=1}^n \Big|nx_i(k_0)-\sum_{j=1}^nx_j(k_0)\Big|\Big)^2 \nonumber\\
&\leq \frac{1}{n^2}\Big(\sum_{i=1}^n \sum_{j=1,\ j\neq i} \Big|x_i(k_0)-x_j(k_0)\Big|\Big)^2 \nonumber\\
&\leq\frac{n-1}{n}\mathcal{H}^2(k_0).
\end{align}

With (\ref{7}) and (\ref{8}) and applying Markov's inequality, we have
\begin{align}\label{11}
\mathbf{P}\Big(\frac{\mathcal{H}(k)}{\mathcal{H}(k_0)}\geq \epsilon\Big)&=\mathbf{P}\Big(\frac{\mathcal{H}^2(k)}{\mathcal{H}^2(k_0)}\geq\epsilon^2\Big)\nonumber\\
&\leq \mathbf{P}\Big(\frac{L(k)}{L(k_0)}\geq  \frac{n}{2(n-1)} \epsilon^2\Big)\nonumber\\
&\leq \frac{2(n-1)}{n} \epsilon^{-2} \frac{\mathbf{E}(L(k))}{L(k_0)}\nonumber\\
&\leq  \frac{2(n-1)}{n} \epsilon^{-2} \prod_{i=k_0}^{k-1}\Big(1-\frac{\lambda_2^\ast}{2n}{P}_i\Big)
\end{align}
where the last inequality holds from (\ref{10}). Since $\sum_{k=m}^{m+T_\ast-1} {P}_k\geq p_\ast$ for all $m\geq0$, according to the arithmetic mean-geometric mean inequality, we have that for all $m\geq0$,
\begin{align}
\prod_{k=m}^{m+T_\ast-1}\Big(1-\frac{\lambda_2^\ast}{2n}P_k\Big)\leq \Big(\frac{T_\ast-\frac{\lambda_2^\ast}{2n}\sum_{k=m}^{m+T_\ast-1}P_k}{T_\ast}\Big)^{T_\ast}\leq\Big(1-\frac{\lambda_2^\ast p_\ast}{2nT_\ast}\Big)^{T_\ast}\doteq c_\ast<1.
\end{align}
As a result, we obtain
\begin{align}\label{12}
\prod_{i=k_0}^{k-1}\Big(1-\frac{\lambda_2^\ast}{2n}{P}_i\Big)\leq c_\ast^{\lfloor\frac{k-k_0}{T_\ast}\rfloor}\leq c_\ast^{\frac{k-k_0}{T_\ast}-1},
\end{align}
where $\lfloor z \rfloor$ denotes the largest integer no larger than $z$.

Then (\ref{11}) and (\ref{12}) lead to
\begin{align}
\mathbf{P}\Big(\frac{\mathcal{H}(k)}{\mathcal{H}(k_0)}\geq \epsilon\Big)\leq  \frac{2(n-1)}{n} \epsilon^{-2}c_\ast^{\frac{k-k_0}{T_\ast}-1},
\end{align}
which implies
\begin{align}
T_{\rm com}(\epsilon)\leq  T_\ast\Big[\frac{3\log \epsilon^{-1} +\log \frac{2(n-1)}{c_\ast n}}{\log c_\ast^{-1}}\Big]={3}\Big[{\log \big(1-\frac{\lambda_2^\ast p_\ast}{2nT_\ast}\big)^{-1}}\Big]^{-1}\log \epsilon^{-1}+O(1).
\end{align}
The desired conclusion follows.

\vspace{3mm}

\noindent (Necessity.) We prove the necessity part of Theorem \ref{thm2} by a contradiction argument. Let $\alpha_1, \alpha_2$ be defined as in the proof of Theorem \ref{thm1}. Set $x_{\alpha_1}(k_0)=0$, $x_{\alpha_2}(k_0)=1$ and $x_{j}(k_0)\in[0,1]$ for all other nodes. Then according to (\ref{15}), we have
\begin{align}\label{16}
\mathbf{P}\Big(\frac{\mathcal{H}(k)}{\mathcal{H}(k_0)}\geq \epsilon\Big)\geq \mathbf{P}\big(x_{\alpha_i}(t)=x_{\alpha_i}(k_0), \ i=1,2; k_0\leq t \leq k\big)=\prod_{t=k_0}^{k-1} \big(1- h_{\alpha_1} {P}_t\big)\big(1- h_{\alpha_2} {P}_t\big)
\end{align}

Take $\epsilon ={1}/{\ell}$ with $\ell=1,2,\dots$. Suppose suitable $p_\ast$ and $T_\ast$ cannot be found such that  $\sum_{k=m}^{m+T_\ast-1} {P}_k\geq p_\ast$ for all $m\geq0$. Then for any $\hat{T}=\ell\log \ell$, there exists an integer  $\hat{k}\geq 0$ such that $\sum_{t=\hat{k}}^{\hat{k}+\hat{T}}{P}_t< 1/2$. According to (\ref{16}) and Lemma \ref{lem2}, we have
\begin{align}
\mathbf{P}\Big(\frac{\mathcal{H}(\hat{k}+\hat{T}+1)}{\mathcal{H}(\hat{k})}\geq \epsilon\Big)&\geq \prod_{t=\hat{k}}^{\hat{k}+\hat{T}} \big(1- h_{\alpha_1} {P}_t\big)\big(1- h_{\alpha_2} {P}_t\big)\nonumber\\
&=e^{\sum_{t=\hat{k}}^{\hat{k}+\hat{T}}\big(\log(1- h_{\alpha_1} {P}_t)+\log(1- h_{\alpha_2} {P}_t)\big)}\nonumber\\
&\geq e^{-2( h_{\alpha_1}+ h_{\alpha_2})\sum_{t=\hat{k}}^{\hat{k}+\hat{T}}{P}_t}\nonumber\\
&> e^{-( h_{\alpha_1}+ h_{\alpha_2})}\nonumber\\
&\geq 1/\ell
\end{align}
for all $\ell\geq e^{ h_{\alpha_1}+ h_{\alpha_2}}$. This immediately implies $T_{\rm com}(1/\ell)\geq \ell\log \ell$, which suggests that $T_{\rm com}(\epsilon)=O(\log\epsilon^{-1})$ does not hold. The proof has been completed.
\section{Convergence under  Independent Communication}
In this section, we focus on the case when the communication between nodes $i$ and $j$ is independent,  as described in the following assumption.

\vspace{3mm}

  \noindent {\bf A3.} {\it (Independent Communication)} The events $\mathbb{E}^+_k $ and $\mathbb{E}^-_k $ are independent for all $k$.
\vspace{3mm}

\begin{remark}
Symmetric and asymmetric randomized gossip algorithms were studied in \cite{f-z}. The symmetry in \cite{f-z} is a deterministic concept where the gossip algorithm is either symmetric or asymmetric, and therefore, it is binary.  While the dependence discussed in this paper carefully characterizes how much symmetry is missing from a probabilistic viewpoint. The symmetric model in \cite{f-z}  is  a special case of ours when $P_k^+=P_k^-=1$ for all $k$ (in this case perfectly dependent communication coincides with independent communication). The asymmetric model in \cite{f-z}  is  a special case of our independent communication model with $P_k^+=1$ and $P_k^-=0$ for all $k$.
\end{remark}

 With A3, $\mathbb{E}^+_0,\mathbb{E}^-_0,\mathbb{E}^+_1,\mathbb{E}^-_1,\dots$ is a sequence of independent events, and the considered gossip algorithm can be expressed as
 \begin{align}\label{20}
 x(k+1)=W(k)x(k),
 \end{align}
where $W(k)$ is the random matrix satisfying

\begin{itemize}
\item[(i)] $\mathbf{P}\Big(W(k)=I-\frac{e_i(e_i-e_j)^T}{2}\Big)=\frac{a_{ij}}{n}{P}_k^+(1-{P}_k^-)+\frac{a_{ji}}{n}{P}_k^-(1-{P}_k^+)$, \ \ $i\neq j$;
\item[(ii)] $\mathbf{P}\Big(W(k)=I-\frac{(e_i-e_j)(e_i-e_j)^T}{2}\Big)=\frac{a_{ij}+a_{ji}}{n}{P}_k^+{P}_k^-$, \ \ $i> j$;
\item[(iii)] $\mathbf{P}\Big(W(k)=I\Big)=1-\sum_{i>j}\frac{a_{ij}+a_{ji}}{n}\Big( {P}_k^++{P}_k^--{P}_k^+{P}_k^-\Big)$.
\end{itemize}

In order to establish the convergence results under independent communication, we need the following condition for the underlying connectivity.

\vspace{3mm}

  \noindent {\bf A4.} {\em (Double Connectivity)} Both the underlying graph $\mathcal {G}_0$ and its converse graph $\mathcal{G}_0^T$ are quasi-strongly connected.

  \vspace{3mm}

  \begin{remark}
  Note that the condition of  $\mathcal {G}_0$ being strongly connected implies A4, but not vice versa. Moreover, it is not hard to see that $\mathcal {G}_0=\mathcal {G}_A$, and $\mathcal{G}_{0}^T=\mathcal {G}_{A^T}$, where $\mathcal {G}_A$ and $\mathcal {G}_{A^T}$ are the induced graph of $A$ and $A^T$, respectively.
  \end{remark}

We now present the main result on a.s. consensus  under independent  communication.
\begin{thm} \label{thm4}
Suppose   A3 (Independent Communication) and A4 (Double Connectivity) hold. Global a.s. {\it consensus}  is achieved  if and only if $\sum_{k=0}^\infty ({P}_k^++{P}_k^-)=\infty$.
\end{thm}

Denote $d_\ast=\max\big\{\rm{diam}(\mathcal{G}_A), \rm{diam}(\mathcal{G}_{A^T})\big\}$, where $\rm{diam}(\mathcal{G}_A)$ and $\rm{diam}(\mathcal{G}_{A^T})$ represent the diameter of the induced graph of $A$ and $A^T$, respectively. Take $E_\ast=|\mathcal{E}_0|-\sum_{i=1}^n \rm{sgn}$$(a_{ii})$, where $|\mathcal{E}_0|$ represents the number of elements in $\mathcal{E}_0$, and $\rm{sgn}$$(z)$ is the sign function. Introduce $a_\ast=\min\{a_{ij}:\ a_{ij}>0, \ i,j=1,\dots,n,\ i\neq j\}$ as the lower bound of the nonzero and non-diagonal entries in the meeting probability matrix $A$. For a.s. $\epsilon$-consensus, the following conclusion holds.
\begin{thm}\label{thm5}
Suppose   A3 (Independent Communication) and A4 (Double Connectivity) hold. Global a.s. {\it $\epsilon$-consensus}  is achieved  if and only if there exist a constant $p_\ast>0$ and an integer $T_\ast\geq1$ such that $\sum_{k=m}^{s+T_\ast-1} ({P}_k^++{P}_k^-)\geq p_\ast$ for all $m\geq0$. In this case, we have
\begin{align}
T_{\rm com}(\epsilon)\leq \frac{4T_\ast\theta_0/p_\ast}{\log\big( 1- \big(\frac{a_\ast}{4n}\big)^{\theta_0} \big)^{-1}} \log \epsilon^{-1} +O(1),
\end{align}
where $\theta_0\doteq(2d_\ast-1)(2E_\ast-1)$.
\end{thm}

We also have the following conclusion indicating that the expected value of the consensus limit $\xi$ equals the initial average $x_{\rm ave}$ if ${P}_k^+={P}_k^-$.
\begin{thm}\label{thm6}
Suppose A3 (Independent Communication) holds and  the consensus limit exists. Then $\mathbf{E}(\xi)=x_{\rm ave}$ if ${P}_k^+={P}_k^-$ for all $k$.
\end{thm}

With independent communication,  when ${P}_k^+={P}_k^-$ for all $k$, it is not hard to see that for any $k=0,1,\dots$, $\mathbf{E}\big(W(k)\big)$ is a  doubly stochastic matrix since it is both stochastic and symmetric. Thus,  Theorem \ref{thm6} holds trivially. Furthermore, we have another conclusion showing that whenever the consensus limit $\xi$ exists, the average  can almost never be preserved.
\begin{thm}\label{thm7}
Suppose A3 (Independent Communication) holds and the number of nodes, $n$, is odd. Assume that $P_k^+, P_k^-\in[0,1-\varepsilon]$ for all $k\geq 0$ with $0<\varepsilon<1$ a fixed number. Then for any $k_0\geq 0$ and for almost all initial conditions $x^0=x(k_0)\in\mathds{R}^n$, we have
\begin{align}\label{22}
\mathbf{P}\Big(\sum_{i=1}^n x_{i}(k)=n x_{\rm ave},\ \  k\geq k_0\Big)=0
\end{align}
if the consensus limit $\xi$ exists.
\end{thm}

\begin{remark}
Surprisingly enough Theorem \ref{thm7} relies on the condition that the number of nodes  is odd. This is not conservative in the sense that we can easily find an example with even number of nodes  such that $\mathbf{P}\big(\sum_{i=1}^n x_{i}(k)=n x_{\rm ave},\ k\geq k_0\big)>0$ conditioned the consensus limit exists.
\end{remark}
\begin{remark}
Note that perfectly dependent communication coincides with independent communication when $P_k^+=P_k^-=1$. This is why we need $P_k^+$ and $P_k^-$ to be at a distance from one in the assumption of Theorem \ref{thm7}.
\end{remark}
Regarding the non-conservativeness of A4 (Double Connectivity) to ensure a consensus under independent communication, we have the following conclusion.
\begin{prop}\label{prop1}
Suppose  A3 (Independent Communication) holds. Then the condition $\sum_{k=0}^\infty ({P}_k^++{P}_k^-)=\infty$ always ensures  an a.s. consensus only if A4 (Double Connectivity) holds.
\end{prop}
{\it Proof.} Assume that A4 does not hold. Then either $\mathcal {G}_0$ or $\mathcal{G}_0^T$ is not quasi-strongly connected.

Let us first discuss the case when $\mathcal {G}_0$ is not quasi-strongly connected.  There will be two distinct nodes $i$ and $j$ such that $\bar{\mathcal {V}}_1\cap \bar{\mathcal {V}}_2=\emptyset$, where $\bar{\mathcal {V}}_1=\{\mbox{nodes\ from\ which\ $i$\ is\ reachable\ in}\  \mathcal
{G}_0\}$ and  $\bar{\mathcal {V}}_2=\{\mbox{nodes\ from\ which\ $j$\ is\ reachable\ in}$\ $\mathcal
{G}_0\}$. The definition of the node pair selection process then implies
\begin{align}
\big\{m: \ a_{mk}>0, k\in \bar{\mathcal{V}}_\tau\big\}\subseteq \bar{\mathcal{V}}_\tau,\ \ \tau=1,2.
\end{align}
Take $P^+_k=0$ for all $k$. Then each node in $\bar{\mathcal {V}}_\tau, \tau=1,2$ can only be connected to the nodes in the same subset. Consensus will then fail with probability one even with $\sum_{k=0}^\infty {P}_k^-=\infty$ if we just take $x_{m}(k_0)=0$ for $m\in\bar{\mathcal {V}}_1$ and $x_{m}(k_0)=1$ for $m\in\bar{\mathcal {V}}_2$.

Similar analysis leads to the same conclusion for the  case with $\mathcal {G}_0^T$  not quasi-strongly connected. This completes the proof. \hfill$\square$
\begin{remark}
It was shown in Remark \ref{rem4} that A1 (Weak Connectivity) is a  lower bound for the underlying connectivity ensuring a consensus, and Theorem \ref{thm1} indicated that under A1, the condition $\sum_{k=0}^\infty ({P}_k^++{P}_k^-)=\infty$ ensures an a.s. consensus under independent communication. Therefore, combining Remark \ref{rem4}, Theorem \ref{thm1}, Theorem \ref{thm4}, and Proposition \ref{prop1}, we can conclude that in terms of consensus convergence of the randomized gossip algorithm, A1 (Weak Connectivity)  is critical  for  perfectly dependent communication, as is A4 (Double Connectivity) for independent communication.
\end{remark}
\begin{remark}
For perfectly dependent communication, Theorem \ref{thm3} shows the consensus limit equals the initial average with probability one. While for independent communication, we see from Theorem \ref{thm6} that only the expected value of the consensus limit equals the initial average with an additional condition $P_k^+=P_k^-$.
\end{remark}
\begin{remark}
We see from (\ref{21}) and (\ref{22})  that with odd number of nodes,  average preserving turns from  almost forever (with probability one for all initial conditions) with perfectly dependent communication, to almost never (with probability zero for almost all initial conditions) for the independent case. As has been shown widely in the classical random graph theory \cite{bollobas,er,kumar}, the probability of a random graph to hold a  certain property often jumps from one to zero when a  certain threshold is crossed.  Now we can conclude that communication dependence provides  such a threshold for average preserving of the considered randomized gossip algorithm.
\end{remark}

With independent communication, we have $W(k)=I-\frac{e_i(e_i-e_j)^T}{2}$ with a nontrivial probability  in (\ref{20}). The matrix $I-\frac{e_i(e_i-e_j)^T}{2}$ is neither  doubly stochastic nor symmetric, and it is  not a projection matrix. These properties are critical for the convergence analysis in Theorems \ref{thm1} and \ref{thm2} to stand. Hence,  we need new methods to analyze the convergence property under assumption A3.

The rest of this section is organized as follows. In Subsection 5.1, we establish several useful lemmas for the convergence analysis. Proofs of Theorems \ref{thm4}, \ref{thm5} and \ref{thm7}  are given in Subsection 5.2.

\subsection{Key Lemmas}
In this subsection,  we first establish an important property of the Bernoulli trials defined by the node communication process. Then, we investigate the product of the stochastic matrices  derived from the gossip algorithm.

\subsubsection{Bernoulli  Communication Links}
Define two (independent) sequences of independent Bernoulli trials
\begin{align}
\mathfrak{B}^+_0, \mathfrak{B}^+_1, \mathfrak{B}^+_2, \dots,\nonumber\\
\mathfrak{B}^-_0, \mathfrak{B}^-_1, \mathfrak{B}^-_2, \dots, \nonumber
\end{align}
 such that $\mathbf{P}\big(\mathfrak{B}^+_k=1\big)=P^+_{k}$ and $\mathbf{P}\big(\mathfrak{B}^-_k=1\big)=P^-_{k}$. Then let
 \begin{align}
\mathfrak{B}_0, \mathfrak{B}_1, \mathfrak{B}_2, \dots,\nonumber
\end{align}
denote the independent Bernoulli trials given by $\mathfrak{B}_k=1$ if and only if $\mathfrak{B}^-_k+\mathfrak{B}^+_k\geq 1$.

Clearly $\mathfrak{B}_k=1$ characterizes the event that at least one side of the communication succeeds at time $k$. In order to establish the success times  property of the process of $\{\mathfrak{B}_k\}_0^\infty$, we need the following lemma.
 \begin{lem}\label{lem3}
Suppose $\{b_k\},\{c_k\}$ are two sequences satisfying $b_k, c_k\in[0,1]$, $k =0,1,\dots$. Then $\sum_{k=0}^\infty \big(b_k+c_k-b_kc_k\big)=\infty$ if and only if $\sum_{k=0}^\infty \big(b_k+c_k\big)=\infty$.
\end{lem}
{\it Proof.} It is straightforward to see that
\begin{align}\label{32}
\frac{1}{2}(b_k+c_k)\leq \max\{b_k,c_k\}\leq b_k+c_k-b_kc_k\leq b_k+c_k.
\end{align}
Then the  desired conclusion follows trivially. \hfill$\square$

The following lemma holds on the success times of $\{\mathfrak{B}_k\}_0^\infty$.
\begin{lem}\label{lem4}
$\mathbf{P}\big(\mbox{for all $k_0\geq0$},\  \mathfrak{B}_k=1\  \mbox{for infinitely many $k\geq k_0$} \big)=1$ if and only if $\sum_{k=0}^\infty ({P}_k^++{P}_k^-)=\infty$.
\end{lem}
{\it Proof.} Note that, we have $\mathbf{P}\big(\mathfrak{B}_k=1\big)=1-(1-P^+_{k})(1-P^-_{k})=P^+_{k}+P^-_{k}-P^+_{k}P^-_{k}\doteq z_k$. Then
\begin{align}
\mathbf{P}\big(\mathfrak{B}_k=0, k\geq k_0\big)=\prod_{k=k_0}^\infty (1-z_k),
\end{align}
and the necessity part of the conclusion follows immediately from Lemmas  \ref{lem1} and \ref{lem3}.

On the other hand, since $\sum_{k=0}^\infty ({P}_k^++{P}_k^-)=\infty$ implies $\prod_{k=s}^\infty\big( 1-z_k\big)=0$ for all $s\geq0$, we have
\begin{align}
\mathbf{P}\big(\mbox{$\exists m\geq0$},\  \mathfrak{B}_k=1\  \mbox{for finitely many $k\geq m$} \big)&\leq \sum_{m=0}^\infty\mathbf{P}\big(\mathfrak{B}_k=1\  \mbox{for finitely many $k\geq m$} \big)\nonumber\\
&\leq \sum_{m=0}^\infty \sum_{s= m}^\infty\mathbf{P}\big(\mathfrak{B}_k=0\  \mbox{for $k\geq s$} \big)\nonumber\\
&\leq \sum_{m=0}^\infty \sum_{s= m}^\infty \prod_{k=s}^\infty\big( 1-z_k\big)\nonumber\\
&=0.
\end{align}
This leads to $\mathbf{P}\big(\mbox{for all $k_0\geq0$},\  \mathfrak{B}_k=1\  \mbox{for infinitely many $k\geq k_0$} \big)=1$. We have now completed the proof. \hfill$\square$
\subsubsection{Products of Transition Matrices}
The considered gossip algorithm is determined by the possible samples of the transition matrix $W(k)$. Denote $\mathbb{M}=\mathbb{M}_1\bigcup\mathbb{M}_2$ with
$$
\mathbb{M}_1=\Big\{ I-\frac{e_i(e_i-e_j)^T}{2}:\ \  a_{ij}+a_{ji}>0, \ i,j=1,\dots,n,\  i\neq j\Big\}
$$
and
$$
\mathbb{M}_2=\Big\{ I-\frac{(e_i-e_j)(e_i-e_j)^T}{2}:\  \ a_{ij}+a_{ji}>0,\  i,j=1,\dots,n,\  i>j\Big\}.
$$
Then $\mathbb{M}$ is the set including all samples of $W(k)$ except for the identity matrix $I$. The following lemma holds on the product of matrices from $\mathbb{M}$.
\begin{lem}\label{lem5}
Let $M_k \in \mathbb{M},\  k=1,\dots,N$ be $N\geq1$ matrices in $\mathbb{M}$. Then we have $\big(\bigcup _{i=1}^N \mathcal{G}_{ M_i}\big) \subseteq \mathcal{G}_{M_N\cdots M_1 }$. Moreover,  all nonzero entries of the product $M_N\cdots M_1$ have lower bound $2^{-N}$.
\end{lem}
{\it Proof.} We just prove the case for $N=2$. Then the  conclusion  follows immediately by  induction.

Note that the nonzero  entries of any matrix in  $\mathbb{M}$ is no smaller than $1/2$. Moreover, all matrices in $\mathbb{M}$  have positive  diagonal entries. Therefore, denoting $\bar{m}_{ij}$, $\hat{m}_{ij}$ and  ${m}^\ast_{ij}$ as the $ij$-entries of $M_1$,  $M_2$ and $M_1M_2$, respectively, we have
\begin{equation}
{m}^\ast_{i_1 i_2}=\sum_{j=1}^n\bar{m}_{i_1j}\hat{m}_{ji_2}\geq \bar{m}_{i_1 i_2}/2 + \bar{m}_{i_1 i_1}/2.
\end{equation}
This implies  ${m}^\ast_{i_1 i_2}>0$ as long as at least one of $\bar{m}_{i_1i_2}$ and $\hat{m}_{i_1i_2}$ is non-zero.  Furthermore, if ${m}^\ast_{i_1 i_2}>0$, it is straightforward to see  that ${m}^\ast_{i_1 i_2} \geq {1}/{4}.$ This completes the proof.  \hfill$\square$

Recall that $d_\ast=\max\big\{\rm{diam}(\mathcal{G}_A), \rm{diam}(\mathcal{G}_{A^T})\big\}$, where $\rm{diam}(\mathcal{G}_A)$ and $\rm{diam}(\mathcal{G}_{A^T})$ represent the diameter of the induced graph of $A$ and $A^T$, respectively. We have the following lemma.
\begin{lem}\label{lem6}
Suppose   A4 (Double Connectivity) holds.
Let $M_1, \dots, M_{2d_\ast-1}$ be $2d_\ast-1$ products of some finite matrices in $\mathbb{M}$.  If $\mathcal{G}_A\subseteq \mathcal{G}_{M_k}$ or $\mathcal{G}_{A^T}\subseteq \mathcal{G}_{ M_k}$ holds for any $k=1,\dots,2d_\ast-1$, then $M_{2d_\ast-1}\cdots M_1$ is a  scrambling matrix.
\end{lem}
{\it Proof.} Since for any $k$, we have $\mathcal{G}_A\subseteq \mathcal{G}_{M_k}$ or $\mathcal{G}_{A^T}\subseteq \mathcal{G}_{ M_k}$, there must be one of $\mathcal{G}_A\subseteq \mathcal{G}_{M_k}$ or $\mathcal{G}_{A^T}\subseteq \mathcal{G}_{ M_k}$  happens at least $d_\ast$ times for $k=1,\dots,2d_\ast-1$. Without loss of generality, we just focus on the case that $\mathcal{G}_A\subseteq  \mathcal{G}_{M_k}$ happens at least $d_\ast$ times. Thus, there exist $1\leq k_1<k_2<\dots<k_{d_\ast}\leq 2d_\ast-1$ such that $\mathcal{G}_A\subseteq \mathcal{G}_{M_{k_s}},\ s=1,\dots,d_\ast$. We now separate the product $M_{2d_\ast-1}\cdots M_1$ into the product of $d_\ast$ matrices:
$$
\bar{M}_1=M_{k_1}\cdots M_1; \ \ \ \   \bar{M}_j=M_{k_{j}}\cdots M_{k_{j-1}+1}, j=2,\dots,d_\ast-1;\ \ \ \ \bar{M}_{d_\ast}=M_{2d_\ast-1}\cdots M_{k_{d_\ast-1}+1}.
$$
Then $M_{2d_\ast-1}\cdots M_1=\bar{M}_{d_\ast}\cdots \bar{M}_1$. Since $\mathcal{G}_A\subseteq \mathcal{G}_{M_{k_s}}$ for each $k_s$,  we have $\mathcal{G}_A\subseteq \mathcal{G}_{M_{k_s}}\subseteq \mathcal{G}_{\bar{M}_{s}} , s=1,\dots,d_\ast$ directly from Lemma \ref{lem5}.

Suppose $i_0$ is a center node of $\mathcal{G}_{A}$. Then for any other node $j_0$, there must be a path $i_0\rightarrow j_0$ in $\mathcal{G}_A$ with length no larger than $d_\ast$. We assume $d(i_0,j_0)=d_0\leq d_\ast$. We redenote $j_0$ as $i_{d_0}$, and let $i_0e_0i_1\dots e_{d_0}i_{d_0}$ be a shortest path from $i_0$ to $i_{d_0}$ in the graph $\mathcal{G}_{A}$.

We denote the $ij$-entry of $\bar{M}_s$ as $\bar{M}_{ij}^{\langle s\rangle}$ for  $s=1,\dots,d_\ast$. Also denote the $ij$-entry of $\bar{M}_2\bar{M}_1$ as $\bar{M}_{ij}^{\langle2\rangle\langle1\rangle}$.  The fact that $(i_0,i_1)\in \mathcal{E}_{A}\subseteq \mathcal{E}_{\bar{M}_1}$ immediately implies $\bar{M}_{i_1i_0}^{\langle1\rangle}>0$ according to the definition of induced graph.
%We further conclude from $\bar{M}_{i_1i_0}^{\langle1\rangle}>0$ that  $\bar{M}_{i_1i_0}^{\langle2\rangle\langle1\rangle}>0$ according to Lemma \ref{lem5}.
Similarly we have $\bar{M}_{i_2i_1}^{\langle2\rangle}>0$ because $(i_1, i_2)\in \mathcal{E}_{A}\subseteq \mathcal{E}_{\bar{M}_2}$. Thus, we obtain
      \begin{align}
      \bar{M}_{i_2i_0}^{\langle2\rangle\langle1\rangle}=\sum_{\tau=1}^n \bar{M}_{i_2\tau}^{\langle2\rangle}  \bar{M}_{\tau i_0}^{\langle1\rangle}\geq \bar{M}_{i_2i_1}^{\langle2\rangle}  \bar{M}_{i_1i_0}^{\langle1\rangle} >0.
      \end{align}
Similar analysis can be proceeded until we eventually obtain
\begin{align}\label{36}
\bar{M}_{i_{d_0} i_0}^{\langle d_0\rangle\dots\langle1\rangle}>0
\end{align}
where $\bar{M}_{i_\tau i_0}^{\langle d_0\rangle\dots\langle1\rangle}$ denotes the ${i_\tau i_0}$-entry of $\bar{M}_{d_0}\cdots \bar{M}_1$.

Denote the $\tau i_0$-entry of $M_{2d_\ast-1}\dots M_1$ as $M_{\tau i_0}^\ast$. Noting  that (\ref{36}) holds for arbitrary $i_{d_0}$, we see from Lemma \ref{lem5}  that
\begin{align}
M_{\tau i_0}^\ast=\bar{M}_{\tau i_0}^{\langle d_\ast\rangle\dots\langle1\rangle}>0, \; \tau=1,\dots,n.
\end{align}
Therefore, according to the definition of $\lambda(\cdot)$ in (\ref{origin}), obtain
$$
\lambda\big(M_{2d_\ast-1}\cdots M_1\big)\leq 1- \min_{j=1,\dots,n}M_{j i_0}^\ast<1.
$$
We have now proved the conclusion. \hfill$\square$

We further denote    $\mathbb{M}_2^\ast=\Big\{ I-\frac{(e_i-e_j)(e_i-e_j)^T}{2}:\    i,j=1,\dots,n,\  i>j\Big\}$. The following lemma is on the impossibility of the finite-time convergence for the product of matrices in $\mathbb{M}_2^\ast$.
\begin{lem}\label{lem7} Suppose $n$ is an odd number. Take matrices $M_\tau\in \mathbb{M}^\ast_2, \ \tau=1,\dots,k, k\geq 1$  arbitrarily. Then  we have $\delta(M_k\cdots M_1)>0$, where $\delta(\cdot)$ is defined  in (\ref{origin}).
\end{lem}
{\it Proof.}  We prove the lemma by a contradiction argument. Suppose there exist an integer $k_\ast\geq1$ and $k_\ast$ matrices $M_\tau\in \mathbb{M}_2^\ast, \tau=1,\dots,k_\ast$  satisfying $\delta(M_{k_\ast}\cdots M_1)=0$. Therefore, there exists $\beta\in \mathds{R}^n$ as an $n\times1$ vector such that
\begin{equation}\label{64}
M_{k_\ast}\cdots M_1=\mathbf{1}\beta^T,
\end{equation}
which implies
\begin{equation}\label{r2}
M_{k_\ast}\cdots M_1 y=\mathbf{1}\beta^T y=z_0\mathbf{1}
\end{equation}
 for all $y=(y_1\dots y_n)^T\in\mathds{R}^n$, where $z_0=\beta^T y$ is a scalar. Since $M_\tau\in \mathbb{M}_2^\ast, \tau=1,\dots,k_\ast$, the average of $y_1,\dots,y_n$ is always
 preserved, and therefore, $z_0=\sum_{i=1}^n y_i/n$.

 Let $n=2n_0+1$ since $n$ is an odd number. Take $y_i=0$ for $i=1,\dots,n_0$ and $y_i=2^{k_\ast+1}$ for $i=n_0+1,\dots,2n_0+1$. Then  $z_0=2^{k_\ast+1}(n_0+1)/(2n_0+1)$. On the other hand, it is straightforward to verify that each element in $M_{k_\ast}\cdots M_1 y$ can only be an even number, say, $S_0$. Since $S_0$ is even, we have $S_0=2^a S_\ast$ for some $0\leq a\leq k_\ast$ an integer and $S_\ast$ an odd number. This leads to
 \begin{align}
 2^{k_\ast+1}\frac{n_0+1}{2n_0+1}=2^a S_\ast,
 \end{align}
  which implies
   \begin{align}\label{r3}
 2^{k_\ast+1-a}(n_0+1)=(2n_0+1)S_\ast.
 \end{align}
 Clearly it is impossible for  (\ref{r3}) to hold true since its left-hand side is an even number, while the right-hand side odd. Therefore such $M_\tau\in \mathbb{M}_2^\ast, \tau=1,\dots,k_\ast$ does not exist and the desired conclusion follows. \hfill$\square$

\subsection{Proofs}
We are now in a place to prove Theorems \ref{thm4}, \ref{thm5} and \ref{thm7}.
\subsubsection{Proof of Theorem \ref{thm4} }
(Necessity.)  We have
\begin{align}\label{25}
\mathbf{P}\big(x_i(k+1)=x_i(k)\big)&\geq 1- \sum_{j=1,\ j\neq i}\Big[\mathbf{P}\big(\mbox{pair $(j,i)$ is selected}\big)\cdot \mathbf{P}\big(\mbox{$i$ receives $x_j(k)$}\big)\nonumber\\
&\ \ +\mathbf{P}\big(\mbox{pair $(i,j)$ is selected}\big)\cdot \mathbf{P}\big(\mbox{$i$ receives $x_j(k)$}\big)\Big]\nonumber\\
&=1-{P}_k^+\sum_{j=1,\ j\neq i}^n\frac{a_{ij}}{n}-{P}_k^-\sum_{j=1,\ j\neq i}^n\frac{a_{ji}}{n}\nonumber\\
&\geq 1- h_i \max\big\{{P}_k^+,{P}_k^-\big\}
\end{align}
where $h_i$ is introduced in (\ref{24}). Note that, from (\ref{32}), we know $\sum_{k=0}^\infty ({P}_k^++{P}_k^-)=\infty$ if and only if $\sum_{k=0}^\infty \max\big\{{P}_k^+,{P}_k^-\big\}=\infty$.
Therefore, with (\ref{25}), the necessity part of Theorem \ref{thm4} follows immediately from the same argument as the proof of the the necessity part in Theorem \ref{thm1}.

\vspace{3mm}
\noindent (Sufficiency.)  Recall that the considered gossip algorithm is determined by the random matrix, $W(k)$ in (\ref{20}). Denote the induced (random) graph of $W(k)$ as $\mathcal{G}_{W(k)}=\big(\mathcal{V},\mathcal{E}_{W(k)}\big)$. Global  a.s. consensus is equivalent to  $\mathbf{P}\big( \lim_{k\rightarrow \infty} \delta  (W_k\cdots W_2W_1)=0\big) =1$.

With independent communication, we have
\begin{align}\label{30}
&\ \ \ \ \mathbf{P}\Big( (i,j)\in\mathcal{E}_{W(k)} \Big|\mbox{at least one of $\mathfrak{B}^+_k$ and $\mathfrak{B}^-_k$ succeeds at time $k$}\Big)\nonumber\\
&=\frac{\big(a_{ji}P^+_k+a_{ij}P_k^-\big)/n}{1-\big(1-P_k^+\big)\big(1-P_k^-\big)}\nonumber\\
&=\frac{a_{ji}P^+_k+a_{ij}P_k^-}{n\big(P_k^++P_k^--P_k^+P_k^-\big)}\nonumber\\
&\geq \frac{a_{ji}P^+_k+a_{ij}P_k^-}{n\big(P_k^++P_k^-\big)}
\end{align}
for all $i,j=1,\dots,n,\ i\neq j$. Here without loss of generality, we assume $P_k^++P_k^->0$.

Recall that $a_\ast=\min\{a_{ij}:\ a_{ij}>0, \ i,j=1,\dots,n,\ i\neq j\}$ is the lower bound of the nonzero and non-diagonal entries in the meeting probability matrix $A$. Based on (\ref{30}), there are two cases.
\begin{itemize}
\item[(i)] When $P^+_k\geq P_k^-$, for all $i,j=1,\dots,n,\ i\neq j$ with $a_{ij}>0$, we have
\begin{align}\label{31}
\mathbf{P}\Big( (j,i)\in\mathcal{E}_{W(k)} \Big|\mbox{at least one of $\mathfrak{B}^+_k$ and $\mathfrak{B}^-_k$ succeeds at time $k$}\Big)\geq \frac{a_\ast}{2n}.
\end{align}
\item[(i)] When $P^+_k< P_k^-$, for all $i,j=1,\dots,n,\ i\neq j$ with $a_{ij}>0$, we have
\begin{align}\label{33}
\mathbf{P}\Big( (i,j)\in\mathcal{E}_{W(k)} \Big|\mbox{at least one of $\mathfrak{B}^+_k$ and $\mathfrak{B}^-_k$ succeeds at time $k$}\Big)\geq \frac{a_\ast}{2n}.
\end{align}
\end{itemize}

Now we introduce the Bernoulli (success) sequence of $\mathfrak{B}_0,\mathfrak{B}_1,\mathfrak{B}_2,\dots$ as
$$
 0\leq \zeta_1<\zeta_2<\dots :\ \ \    \mathfrak{B}_{\zeta_m} = 1
$$
where  $\zeta_m$ is the time of the $m$'th success. Since $\sum_{k=0}^\infty\big(P_k^++P_k^-\big)=\infty$, Lemma \ref{lem4} guarantees that $\zeta_m<\infty$ a.s. for all $m=1,2,\dots$. For simplicity, we assume $k_0=0$ in the rest of the proof.

From (\ref{31}) and (\ref{33}), for $\zeta_1<\zeta_2<\dots $, we can take a sequence of arcs $(i_1,j_1),(i_2,j_2),\dots$ such that
 \begin{align}
 \mathbf{P}\Big( (i_m,j_m)\in\mathcal{E}_{W(\zeta_m)}\Big)\geq \frac{a_\ast}{2n},
 \end{align}
 where $i_m\neq j_m$ for all $m=1,2,\dots$ and  either $a_{i_mj_m}>0$ or $a_{j_mi_m}>0$ holds. Moreover, since the node pair selection process and the node communication process are independent for different instances, events $\big\{(i_m,j_m)\in\mathcal{E}_{W(\zeta_m)},m=1,2,\dots\big\}$ are independent.

Recall that $E_\ast$ is the number of non-self-looped arcs in the underlying graph. From the double connectivity assumption A4, it is not hard to see that we can select $(i_1,j_1),(i_2,j_2),\dots, (i_{2E_\ast-1}$, $j_{2E_\ast-1})$ properly such that
$$ \ \  \mathcal{G}_A \subseteq \bigcup_{m=1}^{2E_\ast-1}\big\{(i_m,j_m)\big\} \ \ \ \mbox{or}\ \ \  \mathcal{G}_{A^T} \subseteq \bigcup_{m=1}^{2E_\ast-1}\big\{(i_m,j_m)\big\}
$$
holds. Thus, based on Lemma \ref{lem5}, denoting $Q_1=W_{\zeta_{2E_\ast-1}}\cdots W_{\zeta_1}$, we have $\mathcal{G}_{A} \subseteq \mathcal{G}_{Q_1}$ or $\mathcal{G}_{A^T} \subseteq \mathcal{G}_{Q_1}$.

Similarly, denoting  $Q_\tau=W_{\zeta_{(2E_\ast-1)\tau}}\cdots W_{\zeta_{(2E_\ast-1)(\tau-1)+1}}$ for $\tau=1,2,\dots$, we have $\mathcal{G}_{A} \subseteq \mathcal{G}_{Q_\tau}$ or $\mathcal{G}_{A^T} \subseteq \mathcal{G}_{Q_\tau}$  for all $\tau$. According to Lemma \ref{lem6}, we have
\begin{align}
\mathbf{P}\Big( \lambda\big (Q_{2d_\ast-1}\cdots Q_1\big)<1\Big)\geq \big( \frac{a_\ast}{2n}\big)^{(2d_\ast-1)(2E_\ast-1)}.
\end{align}
Moreover, since $Q_{2d_\ast-1}\dots Q_1$ is a product of $\theta_0=(2d_\ast-1)(2E_\ast-1)$ matrices in $\mathbb{M}$, Lemma \ref{lem5} further ensures
\begin{align}
\mathbf{P}\Big( \lambda\big (Q_{2d_\ast-1}\cdots Q_1\big)<1-2^{-\theta_0}\Big)\geq \big( \frac{a_\ast}{2n}\big)^{\theta_0}.
\end{align}
Continuing the analysis, we know that for all $F_s=Q_{(2d_\ast-1)s}\cdots Q_{(2d_\ast-1)(s-1)+1}$, $s=1,2,\dots$, we have
\begin{align}
\mathbf{P}\Big( \lambda\big (F_s\big)<1-2^{-\theta_0}\Big)\geq \big( \frac{a_\ast}{2n}\big)^{\theta_0}, \ \ s=1,2,\dots,
\end{align}
which implies
\begin{align}\label{60}
\mathbf{E}\Big( \lambda\big (F_s\big)\Big)\leq 1- \Big(\frac{a_\ast}{4n}\Big)^{\theta_0}, \ \ s=1,2,\dots.
\end{align}

With Fatou's lemma and Lemma \ref{lem0}, we finally have
\begin{align}
 \mathbf{E}\Big( \lim_{m\rightarrow \infty} \delta \big (F_m\cdots F_1\big)\Big) \leq  \lim_{m\rightarrow \infty}\mathbf{E}\Big( \delta \big (F_m\cdots F_1\big)\Big)\leq  \lim_{m\rightarrow \infty}\mathbf{E}\Big( \prod_{s=1}^m \lambda \big (F_s\big)\Big)=0,
\end{align}
which implies
\begin{align}\label{50}
 \mathbf{P}\Big( \lim_{m\rightarrow \infty} \delta \big (F_m\cdots F_1\big)=0\Big) =1.
\end{align}
Note that, (\ref{50}) leads to $\mathbf{P}\big( \lim_{k\rightarrow \infty} \delta  (W_k\cdots W_1)=0\big) =1$ since the definition of  $F_s, s=1,2,\dots$ guarantees $W_k=I$ for all $k\notin \{\zeta_1,\zeta_2,\dots\}$. This completes the proof.
\subsubsection{Proof of Theorem \ref{thm5}}
Similar to the proof of Theorem \ref{thm4}, the necessity part of Theorem \ref{thm5} follows from the same argument as the one used in the proof of Theorem \ref{thm2}. Here we just focus on the sufficiency statement of Theorem \ref{thm5}. For simplicity we assume $k_0=0$.

Denote the $ij$-entry of $W(k-1)\cdots W(0)$ as $\Psi_{ij}(k)$. Then for all $i,j$ and $\alpha$, we have
\begin{align}
\big|x_{i}(k)-x_j(k)\big|&=\Big|\sum_{m=1}^n \Psi_{im }(k)x_{m }(0)-\sum_{m=1}^n \Psi_{jm }(k)x_{m }(0)\Big|\nonumber\\
&=\Big|\sum_{m=1}^n \Psi_{im }(k)\big(x_{m }(0)-x_\alpha(0)\big)-\sum_{m=1}^n \Psi_{jm }(k)\big(x_{m }(0)-x_\alpha(0)\big)\Big|\nonumber\\
&\leq\sum_{m=1}^n\big| \Psi_{im }(k)- \Psi_{jm }(k)\big|\cdot \max_{m}\big|x_{m }(0)-x_\alpha(0)\big|\nonumber\\
&\leq n \delta \big(W(k-1)\cdots W(0)\big)\mathcal{H}(0),
\end{align}
which implies
\begin{align}
\mathcal{H}(k)\leq n \delta \big(W(k-1)\cdots W(0)\big)\mathcal{H}(0).
\end{align}

Then we introduce $\rho_k=\sum_{m=0}^{k-1} \mathfrak{B}_m$. From Markov's inequality, we have
\begin{align}\label{61}
\mathbf{P}\Big(\frac{\mathcal{H}(k)}{\mathcal{H}(0)} \geq\epsilon \Big)&\leq\mathbf{P}\Big( \delta \big(W(k-1)\cdots W(0)\big) \geq \frac{\epsilon}{n} \Big)\nonumber\\
&\leq \frac{n}{\epsilon} \mathbf{E}\Big( \delta \big(W(k-1)\cdots W(0)\big) \Big)\nonumber\\
&=\frac{n}{\epsilon} \mathbf{E}\Big( \delta \big(W_{\zeta_{\rho_k}}\cdots W_{\zeta_1}\big) \Big)\nonumber\\
&\leq  \frac{n}{\epsilon} \mathbf{E}\Big( \lambda\big( F_{\lfloor \frac{\rho_k}{\theta_0}\rfloor}\big)\cdots \lambda\big(F_1\big)\Big),
\end{align}
where $\theta_0$ and $F_m, m=1,2,\dots$ are defined in the proof of Theorem \ref{thm4}. Thus, (\ref{60}) and (\ref{61}) lead to
\begin{align}\label{62}
 \mathbf{E}\Big( \lambda\big( F_{\lfloor \frac{\rho_k}{\theta_0}\rfloor}\big)\cdots \lambda\big(F_1\big)\Big)&=\mathbf{E}\Big(\mathbf{E}\Big( \lambda\big( F_{\lfloor \frac{\rho_k}{\theta_0}\rfloor}\big)\cdots \lambda\big(F_1\big)\big|\rho_k\Big)\Big)\nonumber\\
 &= \Big(1- \big(\frac{a_\ast}{4n}\big)^{\theta_0}\Big)^{\mathbf{E} \big( \lfloor \frac{\rho_k}{\theta_0} \rfloor\big)}\nonumber\\
 &\leq\Big(1- \big(\frac{a_\ast}{4n}\big)^{\theta_0}\Big)^{\frac{\mathbf{E} ( \rho_k)}{\theta_0}-1}\nonumber\\
 &= \Big(1- \big(\frac{a_\ast}{4n}\big)^{\theta_0}\Big)^{\frac{\sum_{m=0}^{k-1} \big(P_k^++P_k^--P_k^+P_k^-\big)}{\theta_0}-1}\nonumber\\
 &\leq \Big(1- \big(\frac{a_\ast}{4n}\big)^{\theta_0}\Big)^{\frac{\sum_{m=0}^{k-1} \big(P_k^++P_k^-\big)}{2\theta_0}-1}
\end{align}
since both the node pair selection  process and the node communication process are independent for different instances.

Since there exist a constant $p_\ast>0$ and an integer $T_\ast\geq1$ such that $\sum_{k=m}^{s+T_\ast-1} ({P}_k^++{P}_k^-)\geq p_\ast$ for all $m\geq0$, (\ref{61}) and (\ref{62}) imply
\begin{align}
\mathbf{P}\Big(\frac{\mathcal{H}(k)}{\mathcal{H}(0)} \geq\epsilon \Big)\leq \frac{n}{\epsilon} \Big(1- \big(\frac{a_\ast}{4n}\big)^{\theta_0}\Big)^{\frac{ \lfloor \frac{k}{T_\ast}\rfloor p_\ast}{2\theta_0}-1}
\end{align}
and thus
\begin{align}
T_{\rm com}(\epsilon)\leq \frac{4T_\ast\theta_0/p_\ast}{\log\big( 1- \big(\frac{a_\ast}{4n}\big)^{\theta_0} \big)^{-1}} \log \epsilon^{-1} +O(1).
\end{align}
The desired conclusion follows.
\subsubsection{Proof of Theorem \ref{thm7}}
With independent communication, it follows from Theorem \ref{thm4} that $\sum_{k=0}^\infty \big( P_k^++P_k^-\big)=\infty$ if the consensus limit $\xi$ exists. It is not hard to see that
\begin{align}
\mathbf{P}\big(W(k)\in \mathbb{M}_1\big)&=\Big(1-\frac{\sum_{i=1}^n a_{ii}}{n}\Big)\Big(P^+_k\big(1-P^-_k\big)+P^-_k\big(1-P^+_k\big)\Big)\nonumber\\
&= \Big(1-\frac{\sum_{i=1}^n a_{ii}}{n}\Big) \big(P^+_k+P^-_k-2P^+_kP^-_k\big).
\end{align}
 Since $P_k^+, P_k^-\in[0,1-\varepsilon]$ for all $k\geq 0$ with $0<\varepsilon<1$, we have
 \begin{align}
 P^+_k+P^-_k-2P^+_kP^-_k &=P^+_k+P^-_k-P^+_kP^-_k-P^+_kP^-_k\nonumber\\
 &\geq \max\{P^+_k, P^-_k\} - (1-\varepsilon)\max\{P^+_k, P^-_k\}\nonumber\\
 &=\varepsilon\max\{P^+_k, P^-_k\}\nonumber\\
 &\geq \frac{\varepsilon}{2}\big( P_k^++P_k^-\big).
 \end{align}
 As a result, we have $\sum_{k=0}^\infty \big( P_k^++P_k^--2P_k^+P_k^-\big)=\infty$. By a similar argument as we obtain Lemma \ref{lem4}, for any $k_0\geq0 $, we have
 \begin{align}\label{83}
\mathbf{P}\Big(W(k)\in \mathbb{M}_1\ \mbox{for infinitely many $k$ with $k\geq k_0$}\Big)=1
\end{align}
conditioned that the consensus limit $\xi$ exists.

Then we show that for almost all initial conditions, it is impossible  to generate finite-time convergence along every sample path $\{W^\omega(k)\}_0^\infty$  of the random matrix process $\{W(k)\}_0^\infty$ which satisfies  $W^\omega(k)\in \mathbb{M}_2$ for all $k$. For any $k=1,2,\dots$, we define
$$
\mathcal{I}_k=\big\{x^0 \in \mathds{R}^n: \ \exists z\in\mathds{R}, \Gamma_i \in \mathbb{M}_2, i=1,\dots,k\ \mbox{s.t.}\ \Gamma_k\cdots \Gamma_1x^0= z\mathbf{1} \big\}.
$$
Suppose $\Gamma_1, \dots, \Gamma_k\in \mathbb{M}_2$. We denote
\begin{align}
\Gamma_k\cdots \Gamma_1=\big(\Phi_1 \dots \Phi_n\big)^T
\end{align}
where $\Phi_m$ is the $m$'th column of  $\Gamma_k\cdots \Gamma_1$. With Lemma \ref{lem7}, we know that $\delta(\Gamma_k\cdots \Gamma_1)>0$, which implies $
\bigcap_{i=1}^n \big({\rm span}\{\Phi_i\}\big)^\bot$
is a linear space with dimension  no larger than $n-2$ noticing that $\Gamma_k\cdots \Gamma_1$ is a stochastic matrix. Therefore, since $\Gamma_k\cdots \Gamma_1x^0= z\mathbf{1}$ leads to $\Gamma_k\dots \Gamma_1\big(x^0- z\mathbf{1}\big)=0$, we have
$$
\mathcal{I}_k=\bigcup_{\Gamma_1, \dots, \Gamma_k\in \mathbb{M}_2} \mathds{R}^1\times \mathcal{Y}_{\Gamma_k\cdots \Gamma_1},
$$
where $\mathcal{Y}_{\Gamma_k\cdots \Gamma_1} \doteq \big\{y\in \mathds{R}^n: \Gamma_k\cdots \Gamma_1 y=0\big\}=\bigcap_{i=1}^n \big({\rm span}\{\Phi_i\}\big)^\bot$.
Noting the fact that $\mathbb{M}_2$ is a finite set, we  further conclude that $\mathbf{Me}\big(\mathcal{I}_k\big)=0$, where $\mathbf{Me}(S)$ for $S\subseteq \mathds{R}^n $ denotes the standard Lebesgue measure in $\mathds{R}^n$.  This immediately implies
\begin{align}\label{r8}
\mathbf{Me}\Big(\bigcup_{k=1}^\infty\mathcal{I}_k\Big)=0.
\end{align}

Now we observe that if $W^\omega(k)=I-\frac{e_{u}(e_u-e_v)^T}{2}$ for some $u,v\in\mathcal{V}, u\neq v$ and  $k\geq k_0$, $\sum_{i=1}^n x_i(k) =nx_{\rm ave}$ implies $x_u^\omega(k)=x_v^\omega(k)$. Therefore, in this case $W^\omega(k)$ can be replaced by $I$ without changing the value of $x^\omega(k+1)$.

Since the node pair selection process and the node communication process are independent with the nodes' states, we conclude from (\ref{83}) and (\ref{r8}) that
\begin{align}
&\ \ \ \ \mathbf{P}\Big( \sum_{i=1}^n x_{i}(k)=n x_{\rm ave},\ \  k\geq k_0\ {\rm and\ consensus\ limit\ \xi\ exists}\Big|{\rm  \xi\ exists}\Big)\nonumber\\
&=\mathbf{P}\Big( \sum_{i=1}^n x_{i}(k)=n x_{\rm ave},\ \  k\geq k_0\ {\rm and\ consensus\  achieved\ in\ infinite\ time}\Big|{\rm  \xi\ exists}\Big)\nonumber\\
 &\ \ \ \ \ \ + \mathbf{P}\Big( \sum_{i=1}^n x_{i}(k)=n x_{\rm ave},\ \  k\geq k_0\ {\rm and\ consensus\  achieved\ in\ finite\ time}\Big|{\rm  \xi\ exists}\Big)\nonumber\\
 &\leq\mathbf{P}\Big(  \mbox{$x_{i}(k)= x_{j}(k)$\ whenever}\  W({k})=I-\frac{e_{i}(e_{i}-e_{j})^T}{2} \in \mathbb{M}_1, k\geq k_0\ {\rm and\ }\nonumber\\
 &\ \ \ \ \ \ \ \ \ \ \ \ \ \ \ \ \ \ \mbox{consensus\  achieved\ in\ infinite\ time} \Big|{\rm  \xi\ exists}\Big)\nonumber\\
 &\ \ \ \ \ \ + \sum_{m=0}^\infty\mathbf{P}\Big( W({\tau}) \in \mathbb{M}_2,\ \tau=k_0,\dots,k_0+m\ {\rm s.t.}\ W(k_0+m)\cdots W(k_0)x^0=z\mathbf{1}\Big|{\rm  \xi\ exists}\Big)\nonumber\\
 &=0+\sum_{m=0}^\infty 0\nonumber\\
 &=0
\end{align}
for all $x^0\in \mathds{R}^n$ except for $\bigcup_{k=1}^\infty\mathcal{I}_k$, which is a set with measure zero.

The desired conclusion follows immediately.
\section{Conclusions}
This paper presented new results on the role of unreliable  node communication in the convergence of  randomized gossip algorithms. The model for the random node pair selection process is defined by  a stochastic matrix which characterizes  the  interactions among the nodes in the network.  A pair of nodes meets at a random instance, and two Bernoulli communication links  are then established  between the nodes. Communication on each link succeeds with a time-dependent probability. We presented a series of necessary and sufficient conditions on the success probability sequence  to ensure a.s. consensus or $\epsilon$-consensus under perfectly dependent and independent communication processes, respectively. The results showed that the communication symmetry  is critical for the convergence.

The results are summarized in the following table. We notice the following characteristics:
\begin{figure}[H]
\centerline{\epsfig{figure=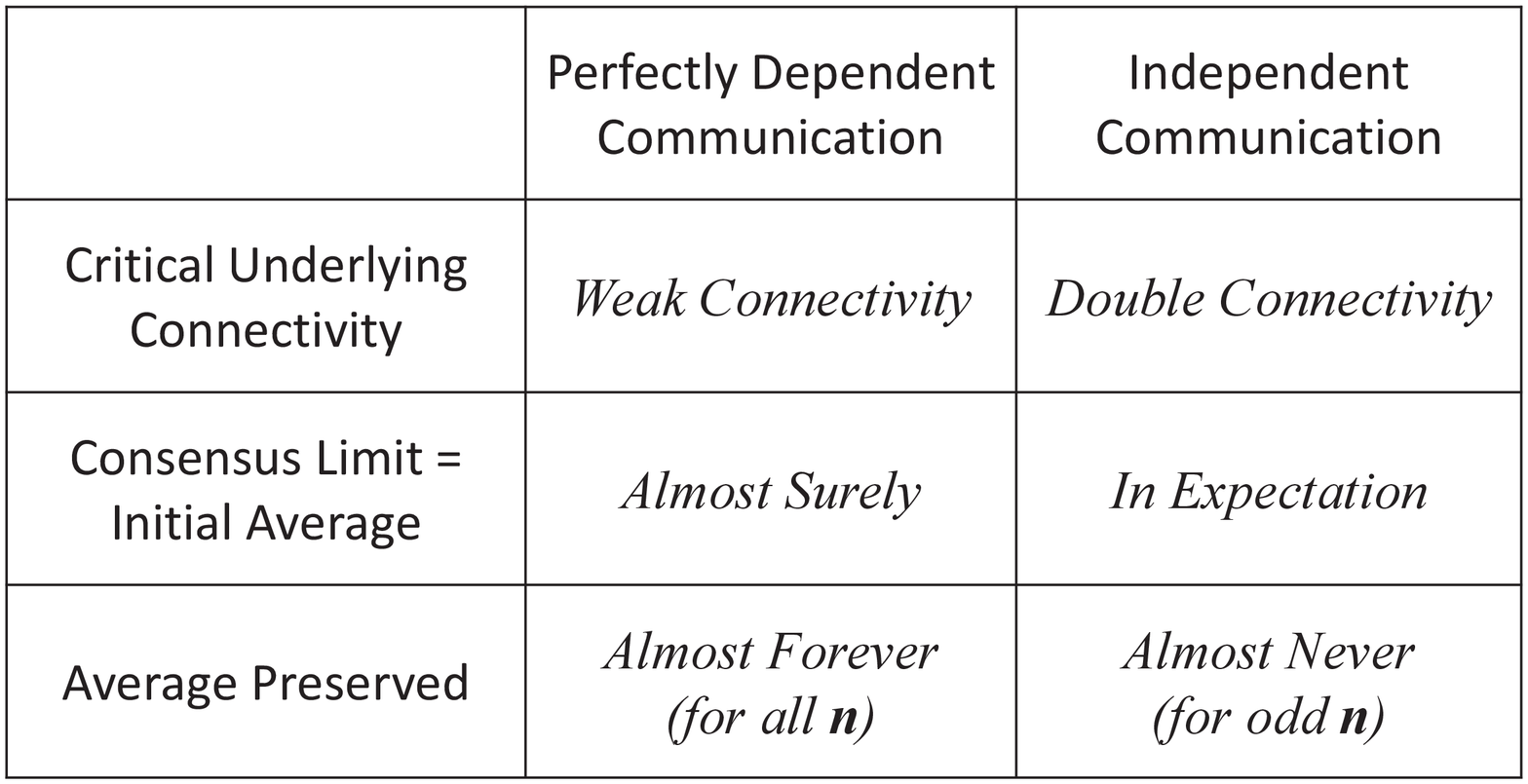, width=0.80\linewidth=0.25}}
\caption{Summary of the properties of the random gossip algorithms considered in the paper. Perfectly dependent  and independent communication gives drastically different behavior. }\label{table}
\end{figure}
\begin{itemize}
\item In terms of consensus convergence of the randomized gossip algorithm, A1 (Weak Connectivity)  is critical  for  perfectly dependent communication, as is A4 (Double Connectivity) for independent communication.

\item For perfectly dependent communication, the consensus limit equals the initial average with probability one. While for independent communication,  only the expected value of the consensus limit equals the initial average for the special case  $P_k^+=P_k^-$.

\item Average is preserved almost forever (with probability one for all initial conditions) with perfectly dependent communication, and it is preserved almost never (with probability zero for almost all initial conditions) with independent communication if the number of nodes is odd.
\end{itemize}

The results illustrate that convergence behavior of distributed algorithms may heavily depend on the probabilistic  dependence properties in the information flow.


\begin{thebibliography}{99}

\bibitem{bert}D. P. Bertsekas and J. N. Tsitsiklis. {\em Introduction to Probability}.  Athena Scientific, Massachusetts, 2002.

\bibitem{lat} G. Latouche, V. Ramaswami. {\em  Introduction to Matrix Analytic Methods in Stochastic Modeling.} 1st edition, ASA SIAM, 1999.


\bibitem{god}
C. Godsil and G. Royle.
\newblock {\em Algebraic Graph Theory.}
\newblock New York: Springer-Verlag, 2001.

\bibitem{ber}
C. Berge and A. Ghouila-Houri.
\newblock{\em Programming, Games, and
Transportation Networks}.
\newblock John Wiley and Sons, New York, 1965.

\bibitem{benv} A. Benveniste, M. M\'{e}tivier and P. Priouret. {\em Adaptive Algorithms and Stochastic Approximations}. Springer-Verlag: Berlin, 1990.

\bibitem{bollobas} B. Bollob\'{a}s. {\em Random Graphs.} Cambridge University Press, second edition, 2001.



\bibitem{kumar} P. Gupta and P. R. Kumar, ``Critical power for asymptotic connectivity in wireless networks,"
{\em Stochastic Analysis, Control, Optimization and Applications: A Volume in Honor of W.H. Fleming}, 547-566, 1998

\bibitem{er}P. Erd\"{o}s and A. R\'{e}nyi, ``On the evolution of random graphs," {\em Publications of the Mathematical Institute of the Hungarian Academy of Sciences}, 17-61, 1960.


\bibitem{gossip2}R. Karp, C. Schindelhauer, S. Shenker, and B. Vöcking, ``Randomized
rumor spreading," in {\em Proc. Symp. Foundations of Computer Science},
pp. 564-574, 2000.

\bibitem{gossip1} D. Kempe, A. Dobra, and J. Gehrke, ``Gossip-based computation of aggregate
information," in {\em Proc. Conf. Foundations of Computer Science},
 pp. 482-491, 2003.



\bibitem{boyd1}S. Boyd,
P. Diaconis and
L. Xiao, ``Fastest mixing markov chain on a graph," {\em SIAM Review}, Vol. 46, No. 4, pp. 667-689, 2004.

\bibitem{boyd} S. Boyd, A. Ghosh, B. Prabhakar and D. Shah, ``Randomized gossip algorithms," {\it IEEE Trans.
Information Theory}, vol. 52, no. 6, pp. 2508-2530, 2006.

\bibitem{}D. Mosk-Aoyama and   D. Shah,  ``Fast distributed algorithms for computing separable functions," {\em IEEE Transactions on Information Theory}, vol.55, no.7, pp. 2997-3007, 2008

    \bibitem{shah}D. Shah, ``Gossip Algorithms," {\em Foundations and Trends in Networking}, Vol. 3, No. 1, pp. 1-125, 2008.


\bibitem{wolf}J. Wolfowitz, ``Products of indecomposable, aperiodic, stochastic matrices,"
{\em Proc. Amer. Math. Soc.}, vol. 15, pp. 733-736, 1963.

\bibitem{haj} J. Hajnal, ``Weak ergodicity in non-homogeneous markov chains," {\em Proc. Cambridge
Philos. Soc.}, no. 54, pp. 233-246, 1958.

\bibitem{degroot} M. H. DeGroot, ``Reaching a consensus," {\it Journal of the American Statistical Association}, vol. 69, no. 345, pp. 118-121, 1974.

\bibitem{cs2} S. Muthukrishnan, B. Ghosh, and M. Schultz, ``First and second order
diffusive methods for rapid, coarse, distributed load balancing,"  {\it Theory
of Computing Systems}, vol. 31, pp. 331-354, 1998.

\bibitem{cs3} R. Diekmann, A. Frommer, and B. Monien, ``Efficient schemes for
nearest neighbor load balancing," {\it Parallel Computing}, vol. 25, pp. 789-812, 1999.

\bibitem{mar}
S. Martinez, J. Cort\'{e}s, and F. Bullo, ``Motion coordination
with distributed information,"
\newblock {\em IEEE Control Systems Magazine}, vol. 27, no. 4, pp. 75-88, 2007.






\bibitem{tsi}
J. Tsitsiklis, D. Bertsekas, and M. Athans, ``Distributed asynchronous
deterministic and stochastic gradient optimization algorithms," {\em
IEEE Trans. Autom. Control}, vol. 31, pp. 803-812, 1986.

\bibitem{jad03}
A. Jadbabaie, J. Lin, and A. S. Morse,
``Coordination of groups of mobile autonomous agents using nearest neighbor rules,"  {\em IEEE Trans. Autom.Control}, vol. 48, no. 6, pp. 988-1001, 2003.


\bibitem{saber04}
R. Olfati-Saber and R. Murray, ``Consensus problems in the networks of agents with switching topology
and time dealys,"
\newblock {\em IEEE Trans. Autom.
Control}, vol. 49, no. 9, pp. 1520-1533, 2004.


\bibitem{fax} J. Fax and R. Murray, ``Information flow and cooperative control
of vehicle formations," {\em IEEE Trans. Autom.
Control}, vol. 49, no. 9, pp. 1465-1476, 2004.

\bibitem{caoming1} M. Cao,  A. S. Morse and B. D. O. Anderson, ``Reaching a consensus in a dynamically changing
environment: a graphical approach," \newblock {\em SIAM J. Control Optim.}, vol. 47, no. 2, 575-600, 2008.

\bibitem{caoming2} M. Cao,  A. S. Morse and B. D. O. Anderson, ``Reaching a consensus in a dynamically changing
environment: convergence rates, measurement
delays, and asynchronous events," \newblock {\em SIAM J. Control Optim.}, vol. 47, no. 2, 601-623, 2008.

\bibitem{caoming3}  M. Cao,  A. S. Morse and B. D. O. Anderson, ``Agreeing asynchronously," {\em IEEE Trans. Autom. Control}, vol. 53, no. 8, 1826-1838, 2008.

\bibitem{ren} W. Ren and R. Beard, ``Consensus seeking in multi-agent systems under dynamically changing interaction topologies," {\em IEEE Trans. Autom. Control}, vol. 50, no. 5, pp. 655-661, 2005.


\bibitem{mor}
L. Moreau, ``Stability of multi-agent systems with time-dependent
communication links," {\em IEEE Trans. Autom. Control}, vol. 50,
pp. 169-182, 2005.



\bibitem{hatano} Y. Hatano and M. Mesbahi, ``Agreement over random networks,"
{\em IEEE Trans. on Autom. Control}, vol. 50, no. 11, pp. 1867-1872,
2005.

\bibitem{wu} C. W. Wu, ``Synchronization and convergence of linear dynamics in random
directed networks," {\em IEEE Trans. Autom. Control}, vol. 51, no. 7,
pp. 1207-1210,  2006.

\bibitem{jad2} A. Tahbaz-Salehi and A. Jadbabaie, ``A necessary and sufficient condition for consensus over
random networks," {\em IEEE Trans. on Autom. Control}, VOL. 53, NO. 3, pp. 791-795, 2008.

\bibitem{f-z}F. Fagnani and S. Zampieri, ``Asymmetric randomized gossip algorithms for consensus," {\em IFAC World Congress}, Seoul, pp. 9051-9056, 2008.

\bibitem{fagnani1} F. Fagnani and S. Zampieri, ``Randomized consensus algorithms over large scale networks," {\it IEEE J. on Selected Areas of Communications}, vol. 26, no.4, pp. 634-649, 2008.

\bibitem{fagnani2}F. Fagnani and S. Zampieri, ``Average consensus with packet drop communication," {\em SIAM J. Control Optim.}, vol. 48, no. 1, pp. 102-133, 2009.




\bibitem{bamieh}S. Patterson, B. Bamieh and A. El Abbadi, ``Convergence rates of distributed average
consensus with stochastic link failures," {\it IEEE Trans.
Autom. Control}, vol. 55, no. 4, pp. 880-892, 2010.

\bibitem{markov}I. Matei, N. Martins and J. S. Baras, ``Almost sure convergence to consensus in markovian random graphs," in {\em Proc. IEEE Conf. Decision and Control}, pp. 3535-3540, 2008.

    \bibitem{roy} C. C. Moallemi and B. Van Roy, ``Consensus propagation," {\it IEEE Trans.
Information Theory}, vol. 52, no. 11, pp. 4753-4766, 2006.

%\bibitem{lifted} K. Jung, D. Shah, and J. Shin, ``Distributed Averaging Via Lifted Markov Chains," {\it IEEE Trans.
%Information Theory}, vol. 56, no. 1, pp. 634-647, 2010.

\bibitem{aysal} T. C. Aysal and K. E. Barner, ``Convergence of consensus models with
stochastic disturbances,"  {\it IEEE Trans.
Information Theory}, vol. 56, no. 8, pp. 4101-4113, 2010.

\bibitem{moura1} S. Kar and J. M. F. Moura, ``Distributed consensus algorithms in sensor networks: quantized data and random link failures," {\em IEEE Trans. signal Processing}, Vol. 58:3, pp. 1383-1400, 2010.

    \bibitem{moura3} U. A. Khan, S. Kar, and J. M. F. Moura, ``Distributed sensor localization in random environments using minimal number of anchor nodes," {\em IEEE Trans. signal Processing}, 57: 5, pp. 2000-2016, 2009.

\bibitem{tsi2} A. Nedi\'{c}, A. Olshevsky, A. Ozdaglar, and J. N. Tsitsiklis, ``On distributed
averaging algorithms and qantization effects," {\it IEEE Trans.
Autom. Control}, vol. 54, no. 11, pp.  2506-2517, 2009.

\bibitem{daron} D. Acemoglu, A. Ozdaglar and A. ParandehGheibi, ``Spread of (Mis)information in social networks," {\em Games and Economic Behavior}, vol. 70, no. 2, pp. 194-227, 2010.

  \bibitem{como} D. Acemoglu, G. Como, F. Fagnani, A. Ozdaglar, ``Opinion fluctuations and persistent disagreement in social networks," in {\em IEEE Conference on Decision and Control}, pp. 2347-2352, Orlando, 2011.

%\bibitem{jad04} A. Jadbabaie, N. Motee, and M. Barahona, ``On the stability of the Kuramoto model of
%coupled nonlinear oscillators", in {\em Proc. American Control Conference},
%Boston, MA, pp. 4296-4301, 2004.

%\bibitem{str} S. H. Strogatz, ``From Kuramoto to Crawford: Exploring the Onset of Synchronization in
%Populations of Coupled Oscillators," {\em Phys. D}, 143, pp. 1–20, 2000.




\end{thebibliography}
\end{document}